\documentclass{aa}
\usepackage{graphicx}
\begin{document}

\title{Cepheid distances from infrared long-baseline interferometry}
\subtitle{I. VINCI/VLTI observations of seven Galactic Cepheids}
\titlerunning{VINCI/VLTI interferometric observations of Cepheids}
\authorrunning{P. Kervella et al.}
\author{P. Kervella\inst{1},
N. Nardetto\inst{2},
D. Bersier\inst{3},
D. Mourard\inst{2}
\and
V. Coud\'e du Foresto\inst{4}
}
\offprints{P. Kervella}
\institute{European Southern Observatory, Alonso de Cordova 3107, Casilla 19001, Vitacura, Santiago 19, Chile
\and D\'epartement Fresnel, UMR CNRS 6528, Observatoire de la C\^ote d'Azur,
   BP 4229, 06304 Nice Cedex 4, France
\and Harvard-Smithsonian Center for Astrophysics, 60 Garden St., Cambridge, MA 02138, USA
\and LESIA, Observatoire de Paris-Meudon, 5, place Jules Janssen, F-92195 Meudon Cedex, France}
\mail{pkervell@eso.org}
\date{Received ; Accepted }
\abstract{
We report the angular diameter measurements of seven classical Cepheids,
X\,Sgr, $\eta$\,Aql, W\,Sgr, $\zeta$\,Gem, $\beta$\,Dor, Y\,Oph and $\ell$\,Car
that we have obtained with the VINCI instrument, installed at ESO's VLT Interferometer (VLTI).
We also present reprocessed archive data obtained with the FLUOR/IOTA instrument on $\zeta$\,Gem,
in order to improve the phase coverage of our observations.
We obtain average limb darkened angular diameter values of
$\overline{\theta_{\rm LD}}[{\rm X\,Sgr}] = 1.471 \pm 0.033$\,mas, 
$\overline{\theta_{\rm LD}}[\eta\,{\rm Aql}] = 1.839 \pm 0.028$\,mas,
$\overline{\theta_{\rm LD}}[{\rm W\,Sgr}] = 1.312 \pm 0.029$\,mas,
$\overline{\theta_{\rm LD}}[\beta\,{\rm Dor}] = 1.891 \pm 0.024$\,mas,
$\overline{\theta_{\rm LD}}[\zeta\,{\rm Gem}] =1.747 \pm 0.061$\,mas,
$\overline{\theta_{\rm LD}}[{\rm Y\,Oph}] = 1.437 \pm 0.040$\,mas, and
$\overline{\theta_{\rm LD}}[\ell\,{\rm Car}] = 2.988 \pm 0.012$\,mas.
For four of these stars, $\eta$\,Aql, W\,Sgr, $\beta$\,Dor, and $\ell$\,Car,
we detect the pulsational variation of their angular diameter. This enables
us to compute directly their distances, using a modified version of the
Baade-Wesselink method:
$d[\eta\,{\rm Aql}] = 276^{+55}_{-38}$\,pc,
$d[{\rm W\,Sgr}] = 379^{+216}_{-130}$\,pc,
$d[\beta {\rm Dor}] = 345^{+175}_{-80}$\,pc,
$d[\ell\,{\rm Car}] = 603^{+24}_{-19}$\,pc.
The stated error bars are statistical in nature.
Applying a hybrid method, that makes use of the Gieren et al.~(\cite{gieren98})
Period-Radius relation to estimate the linear diameters, we obtain the following distances
(statistical and systematic error bars are mentioned):
$d[{\rm X\,Sgr}] = 324 \pm 7 \pm 17$\,pc,
$d[\eta\,{\rm Aql}] = 264 \pm 4 \pm 14$\,pc,
$d[{\rm W\,Sgr}] = 386 \pm 9 \pm 21$\,pc,
$d[\beta {\rm Dor}] = 326 \pm 4 \pm 19$\,pc,
$d[\zeta\,{\rm Gem}] = 360 \pm 13 \pm 22$\,pc,
$d[{\rm Y\,Oph}] = 648 \pm 17 \pm 47$\,pc,
$d[\ell\,{\rm Car}] = 542 \pm 2 \pm 49$\,pc.
}
\maketitle
\section{Introduction}
For almost a century, Cepheids have occupied a central role in
distance determinations. This is thanks to the existence of the 
Period--Luminosity (P--L) relation, $M = a\,\log P + b$, 
which relates the logarithm of the variability period of a Cepheid
to its absolute mean magnitude.
These stars became even more important
since the {\it Hubble Space Telescope} Key Project on the
extragalactic distance scale (Freedman et al.\,\cite{freedman01})
has totally relied on Cepheids for the calibration of distance
indicators to reach cosmologically significant distances.
In other words, if the calibration of the Cepheid P--L relation is
wrong, the whole extragalactic distance scale is wrong.

There are various ways to calibrate the P--L relation.
The avenue chosen by the $HST$ Key-Project was to \emph{assume} a
distance to the Large Magellanic Cloud (LMC), thereby adopting a zero
point of the distance scale.
Freedman et al.\,(\cite{freedman01}) present an extensive discussion
of all available LMC distances, and note, with other authors
(see e.g. Benedict et al.\,\cite{benedict02}), that the LMC distance is
currently the weak link in the extragalactic distance scale ladder.
Another avenue is to
determine the zero point of the P--L relation with Galactic Cepheids,
using for instance parallax measurements, Cepheids in clusters, or
through the Baade-Wesselink (BW) method. 
We propose in this paper and its sequels (Papers II and III) to improve the
calibration of the Cepheid P--R, P--L and surface brightness--color relations
through a combination of spectroscopic and interferometric observations of
bright Galactic Cepheids.

In the particular case of the P--L relation, the slope $a$
is well known from Magellanic Cloud Cepheids (e.g. Udalski et al.\,\cite{udalski99}),
though Lanoix et al.\,(\cite{lanoix99}) have suggested
that a Malmquist effect (population incompleteness) could bias this value.
On the other hand, the calibration of the zero-point $b$ (the hypothetic absolute
magnitude of a 1-day period Cepheid) requires measurement of the distance to a
number of nearby Cepheids with high precision. For this purpose,
interferometry enables a new version of the Baade-Wesselink method,
(BW, Baade\,\cite{baade26}, Wesselink\,\cite{wesselink46})
for which we do not need to measure the star's temperature, as we have directly
access to its angular diameter (Davis\,\cite{davis79}; Sasselov \& Karovska\,\cite{sasselov94}).
Using this method, we derive directly the distances to the four nearby Cepheids
$\eta$\,Aql, W\,Sgr, $\beta$\,Dor and $\ell$\,Car.
For the remaining three objects of our sample, X\,Sgr, $\zeta$\,Gem and Y\,Oph,
we apply a hybrid method to derive their distances, based on published values of
their linear diameters.

After a short description of the VINCI/VLTI instrument (Sect.\,\ref{inst_setup}),
we describe the sample Cepheids that we selected (Sect.\,\ref{sample_cephs}).
In Sect.~\ref{data_proc} and \ref{data_quality}, we report our new observations
as well as reprocessed measurements of $\zeta$\,Gem
retrieved from the FLUOR/IOTA instrument archive.
Sect.~\ref{angdiam} is dedicated to the computation of the corresponding
angular diameter values, taking into account
the limb darkening and the bandwidth smearing effects.
In Sect.~\ref{radius_curves_sect} and \ref{puls_param}, we investigate
the application of the BW method to our data, and we
derive the Cepheid distances.

We will discuss the consequences of these results for the calibration of the
Period-Radius (P--R), Period-Luminosity (P--L) and Barnes-Evans relations
of the Cepheids in forthcoming papers (Paper II and III).

\section{Instrumental setup \label{inst_setup}}
\begin{figure}[t]
\centering
\includegraphics[width=8.5cm]{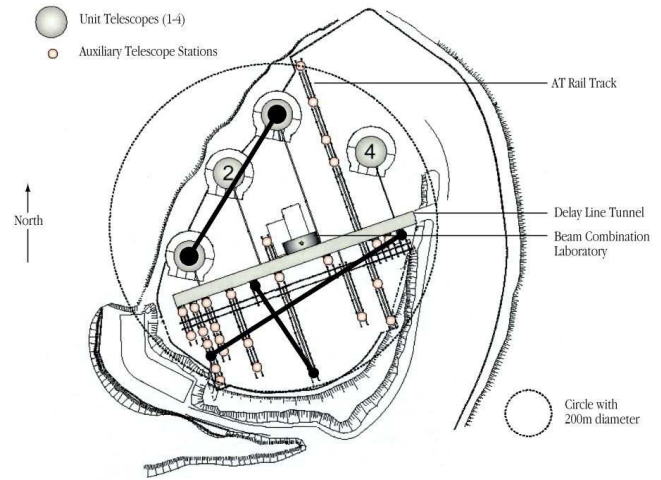}
\caption{Layout of the three baselines used for the VINCI/VLTI Cepheids observations,
UT1-UT3 (102.5\,m), E0-G1 (66\,m) and B3-M0 (140\,m).}
\label{vlti_platform}
\end{figure}

The European Southern Observatory's Very Large Telescope Interferometer
(Glindemann et al. \cite{glindemann}) is in operation on Cerro Paranal, in Northern Chile since
March 2001. For the observations reported in this paper, the beams from two Test Siderostats
(0.35\,m aperture) or two Unit Telescopes (8\,m aperture) were recombined coherently in VINCI,
the VLT INterferometer Commissioning Instrument (Kervella et al. \cite{kervella00}, \cite{kervella03a}). 
We used a regular K band filter ($\lambda = 2.0-2.4\,\mu$m) that gives an effective observation
wavelength of $2.18\,\mu$m for the effective temperature of typical Cepheids
(see Sect.\,\ref{smearing} for details).
Three VLTI baselines were used for this program: E0-G1, B3-M0 and
UT1-UT3 respectively 66, 140 and 102.5\,m in ground length.
Fig.\,\ref{vlti_platform} shows their positions on the VLTI platform.

%______________ Figure
%
%_______________________________________________________ Section
%
\section{Selected sample of Cepheids \label{sample_cephs}}
%____________________Table
\begin{table*}
\caption[]{Relevant parameters of the observed sample of Cepheids, sorted by increasing period.
\label{cepheids_params}}
\begin{tabular}{lccccccc}
\hline
\noalign{\smallskip}

 & X~Sgr & $\eta$~Aql & W~Sgr & $\beta$~Dor & $\zeta$~Gem & Y~Oph & $\ell$~Car \\
  & \object{HD 161592} & \object{HD 187929} & \object{HD 164975} & \object{HD 37350} &
\object{HD 52973} & \object{HD 162714} & \object{HD 84810}\\
\hline

\noalign{\smallskip}

${m_\mathrm{V}}^{\mathrm{a}}$ & 4.581 & 3.942 & 4.700 & 3.731 & 3.928 & 6.164 & 3.771 \\
${m_\mathrm{K}}^{\mathrm{b}}$ & 2.56 & 1.966 & 2.82 & 1.959 & 2.11 & 2.682 & 1.091 \\
Sp. Type & F5-G2II & F6Ib-G4Ib & F4-G2Ib & F4-G4Ia-II & F7Ib-G3Ib & F8Ib-G3Ib & F6Ib-K0Ib\\
%Classification$^{\mathrm{***}}$ & & C & C & & s & & \\
$\pi$ (mas)$^{\mathrm{c}}$ & $3.03 \pm 0.94$ & $2.78 \pm 0.91$ & $1.57 \pm 0.93$ & $3.14 \pm 0.59$ &
$2.79 \pm 0.81$ & $1.14 \pm 0.80$ & $2.16 \pm 0.47$\\
\noalign{\smallskip}
\hline
\noalign{\smallskip}
Min $\rm T_{\mathrm{eff}}$ (K) & 5670 & 5400 & 5355 & 5025 & 5150 & &  \\
Mean $\rm T_{\mathrm{eff}}$ (K)$^{\mathrm{d}}$ & 6150 & 5870 & 5769 & 5490 & 5430 & {\it 5300} & 5090 \\
Max $\rm T_{\mathrm{eff}}$ (K) & 6820 & 6540 & 6324 & 6090 & 5750 & &  \\
Min $\log g$ & 1.86 & 1.25 & 1.72 & 1.60 \\
Mean $\log g ^{\mathrm{e}}$ & 2.14 & 1.49 & 1.82 & 1.83 & 1.50 & {\it 1.50} & 1.50 \\
Max $\log g$ & 2.43 & 1.73 & 2.02 & 2.06 \\
$[\mathrm{M}/\mathrm{H}]^{\mathrm{e}}$ & 0.04 & 0.05 & -0.01 & -0.01 & 0.04 & 0.05 & 0.30 \\
\noalign{\smallskip}

\hline
\noalign{\smallskip}

$T_0$ (JD-$2.452.10^6$)$^{\mathrm{f}}$ & 723.9488 & 519.2477 & 726.8098 & 214.2153 & 210.7407 & 715.4809 & 290.4158 \\
$P$ (days)$^{\mathrm{g}}$ & 7.013059 & 7.176769 & 7.594904 & 9.842425 & 10.150967 & 17.126908 & 35.551341 \\
\noalign{\smallskip}

\hline
Intensity profiles$^{\mathrm{h}}$\\

$a_1$ & $+0.7594$ & $+0.8816$ & $+0.8002$ & $+0.7969$ & $+0.8713$ & $+0.8549$ & $+0.8500$ \\
$a_2$ & $-0.4530$ & $-0.7418$ & $-0.5135$ & $-0.4596$ & $-0.6536$ & $-0.5602$ & $-0.4991$ \\
$a_3$ & $+0.0347$ & $+0.3984$ & $+0.1583$ & $+0.1341$ & $+0.3283$ & $+0.2565$ & $+0.2113$ \\
$a_4$ & $+0.0751$ & $-0.0778$ & $+0.0109$ & $+0.0082$ & $-0.0610$ & $-0.0437$ & $-0.0340$ \\
\noalign{\smallskip}

\hline
\end{tabular}

\begin{list}{}{}
\item[$^{\mathrm{a}}$] $m_V$ from Barnes et al.\,(\cite{barnes87}) for X\,Sgr,
from Barnes et al.\,(\cite{barnes97}) for $\eta$\,Aql, from Moffett \& Barnes\,(\cite{moffett84})
for W\,Sgr and $\zeta$\,Gem, from Berdnikov \& Turner\,(\cite{berdnikov01})
for $\beta$~Dor and $\ell$~Car, and from Coulson \& Caldwell (\cite{coulson85}) for Y\,Oph.
\item[$^{\mathrm{b}}$] $m_K$ from Welch et al.\,(\cite{welch84}) for
X\,Sgr, and W\,Sgr, from Laney \& Stobie\,(\cite{laney92}) for $\beta$\,Dor,
Y\,Oph, and $\ell$\,Car, from Ducati et al.\,(\cite{ducati01}) for $\zeta$\,Gem,
from Barnes et al.\,(\cite{barnes97}) for $\eta$\,Aql.
%\item[$^{\mathrm{c}}$] Classification according to Bersier \& Burki\,(\cite{bersier96}).
\item[$^{\mathrm{c}}$] Parallaxes from the {\sc Hipparcos} catalogue (Perryman et al. \cite{hip}).
\item[$^{\mathrm{d}}$] From Kiss \& Szatm\`ary\,(\cite{kiss98}) for $\zeta$\,Gem and $\eta$\,Aql,
Bersier et al.\,(\cite{bersier97}) for W\,Sgr, and Pel\,(\cite{pel78}) for X\,Sgr and $\beta$\,Dor.
\item[$^{\mathrm{e}}$] From Andrievsky et al.\,(\cite{andrievsky02}), Cayrel de Strobel et al. (\cite{cayrel}, \cite{cayrel01}),
and Pel\,(\cite{pel78}), except for $\log g$ of Y\,Oph.
%from Fernie's\,(\cite{fernie95a}) law.
\item[$^{\mathrm{f}}$] Reference epoch $T_0$ values have been computed near the dates of the VINCI observations,
from the values published by Szabados\,(\cite{szabados89a}).
\item[$^{\mathrm{g}}$] $P$ values from Szabados\,(\cite{szabados89a}).
The periods of $\eta$\,Aql, $\zeta$\,Gem and W\,Sgr are known to evolve.
The values above correspond to the $T_0$ chosen for these stars.
\item[$^{\mathrm{h}}$] Four-parameters intensity profiles from Claret\,(\cite{claret00}) in the $K$ band,
assuming a microturbulence velocity of 4\,km/s and the average values of $\rm T_{\rm eff}$ and $\log g$.
\end{list}

\end{table*}
Despite their brightness, Cepheids are located at large distances, and the {\sc Hipparcos}
satellite (Perryman et al.\,\cite{hip}) could only obtain a limited number of Cepheid
distances with a relatively poor precision.
If we exclude the peculiar first overtone Cepheid $\alpha$\,UMi (Polaris), the closest Cepheid
is $\delta$\,Cep, located at  approximately 250\,pc (Mourard et al.\,\cite{mourard97},
Nordgren et al.\,\cite{nordgren00}).
As described by Davis\,(\cite{davis79}) and Sasselov \& Karovska\,(\cite{sasselov94}),
it is possible to derive directly the distance to the Cepheids for which we can measure the amplitude
of the angular diameter variation. Even for the nearby Cepheids, this requires
an extremely high resolving power, as the largest Cepheid in the sky, $\ell$\,Car,
is only $0.003"$ in angular diameter. Long baseline interferometry is therefore
the only technique that allows us to resolve these objects.
As a remark, the medium to long period Cepheids ($D \approx 200$\,D$_\odot$) in the Large Magellanic
Cloud (LMC) ($d \approx 55$\,kpc) are so small ($\theta \approx 30\,\mu$as)
that they would require a baseline of 20\,km to be resolved in the $K$ band (5\,km in the visible).
However, such a measurement is highly desirable, as it would provide a
precise geometrical distance to the LMC, a critical step in the
extragalactic distance ladder.

Mourard\,(\cite{mourard96}) has highlighted the capabilities of the VLTI for
the observation of nearby Cepheids, as it provides long baselines (up to 202\,m) and thus
a high resolving power. 
Though they are supergiant stars, the Cepheids are
very small objects in terms of angular size.
A consequence of this is that the limit on the number of interferometrically
resolvable Cepheids is not set by the size of the light collectors, but
by the baseline length.
From photometry only, several hundred Cepheids can produce interferometric
fringes using the VLTI Auxiliary Telescopes (1.8\,m in diameter). However,
in order to measure accurately their size, one needs to resolve their disk
to a sufficient level.
Kervella\,(\cite{kervella01a}) has compiled a list of
more than 30 Cepheids that can be measured from Paranal using
the VINCI and AMBER (Petrov et al.\,\cite{petrov00}) instruments.
Considering the usual constraints  in terms of sky coverage, limiting magnitude
and accessible resolution, we have selected seven bright Cepheids observable
from Paranal Observatory (latitude $\lambda = -24 \deg$):
X\,Sgr, $\eta$\,Aql, W\,Sgr, $\beta$\,Dor, $\zeta$\,Gem, Y\,Oph and $\ell$~Car.
The periods of these stars cover a wide range, from 7 to 35.5 days. This
coverage is important to properly constrain the P--R and P--L relations. 
To estimate the feasibility of the observations, the angular diameters of these
stars were deduced from the BW studies by Gieren et al.\,(\cite{gieren93}).
For $\zeta$\,Gem and $\eta$\,Aql, previously published direct interferometric
measurements by Nordgren et al.\,(\cite{nordgren00}), Kervella et al.\,(\cite{kervella01b})
and Lane et al.\,(\cite{lane02}) already demonstrated the feasibility of the observations.
The relevant parameters of the seven Cepheids of our sample, taken from the literature,
are listed in Table\,\ref{cepheids_params}.

%
%_______________________________________________________ Section
%
\section{Interferometric data processing \label{data_proc}}
\subsection{Coherence factors}
We used a modified version (Kervella et al.\,\cite{kervella03c}) of the standard VINCI data
reduction pipeline, whose general principle is based on the original algorithm
of the FLUOR instrument (Coud\'e du Foresto et al.\,\cite{cdf97},
Coud\'e du Foresto et al.\,\cite{coude98a}). The VINCI/VLTI commissioning
data we used for this study are publicly available through the ESO Archive,
and result from two proposals of our group, that were accepted
for ESO Periods 70 and 71.

The goal of the raw data processing is to extract the value of the modulated power
contained in the interferometric fringes.
This value is proportional to the squared visibility $V^2$ of the source on
the observation baseline, which is in turn directly linked to the Fourier transform
of the light distribution of the source through the Zernike-Van Cittert theorem.

One of the key advantages of VINCI is to use single-mode fibers to filter out
the perturbations induced by the turbulent atmosphere. The wavefront
that is injected in the fibers is only the mode guided by the
fiber (gaussian in shape, see Ruilier\,\cite{ruilier99}
or Coud\'e du Foresto\,\cite{coude98b} for details).
The atmospherically corrupted part of the wavefront
is not injected into the fibers and is lost into the cladding. 
Due to the temporal fluctuations of the turbulence, the injected flux changes
considerably during an observation. However, VINCI derives two photometric
signals that can be used to subtract the intensity fluctuations from the interferometric
fringes and normalize them continuously. The resulting calibrated interferograms are
practically free of atmospheric corruption, except the piston mode (differential
longitudinal delay of the wavefront between the two apertures) that tends to smear
the fringes and affect their visibility. Its effect is largely diminished by using a sufficiently
high scanning frequency, as was the case for the VINCI observations.

After the photometric calibration has been achieved,
the two interferograms from the two interferometric outputs
of the VINCI beam combiner are subtracted to remove the residual
photometric fluctuations.
As the two fringe patterns are in perfect phase opposition, this subtraction
removes a large part of the correlated fluctuations and enhances the
interferometric fringes.
Instead of the classical Fourier analysis, we
implemented a time-frequency analysis (S\'egransan et al. \cite{s99})
based on the continuous wavelet transform (Farge \cite{farge92}).
In this approach, the projection of the signal is not onto a sine wave (Fourier transform),
but onto a function, i.e. the wavelet, that is localised in both time and frequency.
We used as a basis the Morlet wavelet, a gaussian envelope
multiplied by a sine wave. With the proper choice of the number of oscillations
inside the gaussian envelope, this wavelet closely matches a VINCI
interferogram. It is therefore very efficient at localizing the signal in both time and frequency.

The differential piston corrupts  the amplitude and the shape of the
fringe peak in the wavelet power spectrum. 
A selection based on the shape of fringe peak in the 
time-frequency domain is used to remove ``pistonned'' and false detection interferograms.  
Squared coherence factors $\mu^2$ are then derived by
integrating the wavelet power spectral density (PSD)
of the interferograms at the position and frequency of the fringes.
The residual photon and detector noise backgrounds are removed
by making a least squares fit of the PSD at high and low frequency.

\subsection{Calibrators}

The calibration of the Cepheids' visibilities was achieved using
well-known calibrator stars that have been selected in the
Cohen et al. (\cite{cohen99}) catalogue, with the exception of $\epsilon$\,Ind.
This dwarf star was measured separately (S\'egransan et al.\,\cite{segransan03})
and used to calibrate one of the $\eta$\,Aql measurements.
The angular diameters of 39\,Eri\,A, HR\,4050 and HR\,4546
(which belong to the Cohen et al.\,\cite{cohen99} catalogue)
were also measured separately, as these stars appeared to give
a slightly inconsistent value of the interferometric efficiency.

For 39\,Eri\,A and HR\,4546, the measured angular diameters
we find are $\theta_{\rm UD} = 1.74 \pm 0.03$ and $2.41 \pm 0.04$\,mas, respectively.
These measured values are only $2\,\sigma$ lower than the Cohen et al~(\cite{cohen99})
catalogue values of $\theta_{\rm UD} = 1.81 \pm 0.02$ and $2.53 \pm 0.04$\,mas.
A possible reason for this difference could be the presence of faint,
main sequence companions in orbit around these two giant stars.
The additional contribution of these objects would bias the diameter found
by spectrophotometry towards larger values, an effect consistent with what we observe.
For HR\,4050, we obtained $\theta_{\rm UD} = 5.18 \pm 0.05$\,mas,
only +1\,$\sigma$ away from the catalogue value of $\theta_{\rm UD} = 5.09 \pm 0.06$\,mas, .
The characteristics of the selected calibrators are listed in Table~\ref{calib_params}.
The limb-darkened disk (LD) angular diameters of these stars were converted into uniform disk
values using linear coefficients taken from Claret et al.~(\cite{claret95}). As demonstrated by
Bord\'e et al. (\cite{borde}), the star diameters in the Cohen et al.~(\cite{cohen99})
list have been measured very homogeneously to a relative precision of approximately 1\% and agree
well with other angular diameter estimation methods.

The calibrators were observed soon before and after the Cepheids,
in order to verify that the interferometric efficiency (IE) has not changed
significantly during the Cepheid observation itself. In some cases, and due to the technical
nature of commissioning observations, part of the Cepheid observations could not be
bracketed, but only immediately preceded or followed by a calibrator.
However, the stability of the IE has proved to be generally very good, and we
do not expect any significant bias from these single-calibrator observations.
Some observations included several calibrators to allow a cross-check of
of their angular sizes. The calibrators were chosen as close
as possible in the sky to our target Cepheids, in order to be able
to observe them with similar airmass.
This selection has taken into account the constraints in terms of limiting
magnitude and sky coverage imposed by the VLTI siderostats and delay lines.
The IE was computed from the coherence factor measurements obtained on
the calibrators, taking into account the bandwidth smearing effect (see Sect.~\ref{smearing})
and a uniform disk angular diameter model. This calibration process yielded the final
squared visibilities listed in Tables~\ref{table_angdiams_x_sgr} to \ref{table_angdiams_l_car}.
\begin{table*}
\caption[]{Relevant parameters of the calibrators.\label{calib_params}}
\begin{tabular}{llcclccrll}
\hline
\noalign{\smallskip}
Name & & $m_\mathrm{V}$ & $m_\mathrm{K}$ & Sp.\,Type & $\rm T_{\mathrm{eff}}$(K) & $\log g$ & $\pi$ (mas)$^{\mathrm{a}}$& ${\theta_{\rm {LD}}}$(mas)$^{\mathrm{b}}$ & ${\theta_{\rm {UD}}}$(mas)$^{\mathrm{c}}$ \\
 \noalign{\smallskip}
\hline
\noalign{\smallskip}
$\chi$\,Phe & \object{HD 12524} & 5.16 & 1.52 & K5III & 3780 & 1.9 & $8.76 \pm 0.64$ & $2.77 \pm 0.032$ & $2.69 \pm 0.031$\\
39\,Eri\,A & \object{HD 26846} & 4.90 & 2.25 & K3III & 4210 & 2.2 & $15.80 \pm 0.95$ & $1.79 \pm 0.031^{*}$ & $1.74 \pm 0.030^{*}$ \\
$\epsilon$\,Ret & \object{HD 27442} & 4.44 & 1.97 & K2IVa & 4460 & 2.3 & $54.84 \pm 0.50$ & $1.95\pm 0.049$ & $1.90 \pm 0.048$ \\
HR\,2533 & \object{HD 49968} & 5.69 & 2.10 & K5III & 3780 & 1.9 & $6.36 \pm 0.92$ & $1.93 \pm 0.020$ & $1.87 \pm 0.019$ \\
HR\,2549 & \object{HD 50235} & 5.00 & 1.39 & K5III & 3780 & 1.9 & $3.60 \pm 0.56$ & $2.25 \pm 0.036$ & $2.18 \pm 0.035$ \\
$\gamma^2$\,Vol & \object{HD 55865} & 3.77 & 1.52 & K0III & 4720 & 2.6 & $23.02 \pm 0.69$ & $2.50 \pm 0.060$ & $2.44 \pm 0.059$ \\
6\,Pup & \object{HD 63697} & 5.18 & 2.62 & K3III & 4210 & 2.2 & $12.87 \pm 0.71$ & $1.88 \pm 0.039$ & $1.83 \pm 0.038$\\
HR\,3046 & \object{HD 63744} & 4.70 & 2.31 & K0III & 4720 & 2.6 & $14.36 \pm 0.48$ & $1.67 \pm 0.025$ & $1.63 \pm 0.024$ \\
HR\,4050 & \object{HD 89388} & 3.38 & 0.60 & K3IIa & 4335 & 2.3 & 4.43 $\pm$ 0.49 & $5.32 \pm 0.050^{*}$ & $5.18 \pm 0.048^{*}$\\
HR\,4080 & \object{HD 89998} & 4.83 & 2.40 & K1III & 4580 & 2.5 & $16.26 \pm 0.56$ & $1.72 \pm 0.020$ & $1.68 \pm 0.019$ \\
HR\,4526 & \object{HD 102461} & 5.44 & 1.77 & K5III & 3780 & 1.9 & $3.97 \pm 0.61$ & $3.03 \pm 0.034$ & $2.94 \pm 0.033$ \\
HR\,4546 & \object{HD 102964} & 4.47 & 1.56 & K3III & 4210 & 2.2 & $7.03 \pm 0.72$ & $2.48 \pm 0.036^{*}$ & $2.41 \pm 0.035^{*}$\\
HR\,4831 & \object{HD 110458} & 4.67 & 2.28 & K0III & 4720 & 2.6 & $17.31 \pm 0.65$ & $1.70 \pm 0.018$ & $1.66 \pm 0.018$\\
$\chi$\,Sco & \object{HD 145897} & 5.25 & 1.60 & K3III & 4210 & 2.2 & $7.43 \pm 0.91$ & $2.10 \pm 0.023$ & $2.04 \pm 0.022$ \\
70\,Aql & \object{HD 196321} & 4.90 & 1.21 & K5II & 3780 & 1.9 & $1.48 \pm 0.91$ & $3.27 \pm 0.037$ & $3.17 \pm 0.036$ \\
7\,Aqr & \object{HD 199345} & 5.50 & 2.00 & K5III & 3780 & 1.9 & $5.42 \pm 0.99$ & $2.14 \pm 0.024$ & $2.08 \pm 0.023$ \\
$\epsilon$\,Ind & \object{HD 209100} & 4.69 & 2.18 & K4.5V & 4580 & 4.5 & $275.79 \pm 0.69$ & $1.89 \pm 0.051^{*}$ & $1.84 \pm 0.050^{*}$\\
$\lambda$\,Gru & \object{HD 209688} & 4.48 & 1.68 & K3III & 4210 & 2.2 & $13.20 \pm 0.78$ & $2.71 \pm 0.030$ & $2.64 \pm 0.029$ \\
HR\,8685 & \object{HD 216149} & 5.41 & 1.60 & M0III & 3660 & 1.4 & $2.95 \pm 0.69$ & $2.07 \pm 0.021$ & $2.01 \pm 0.020$ \\
\noalign{\smallskip}
\hline
\end{tabular}
\begin{list}{}{}
\item[$^{\mathrm{a}}$] Parallaxes from the {\sc Hipparcos} catalogue (Perryman et al.\,\cite{hip}).
\item[$^{\mathrm{b}}$] Catalogue values from Cohen et al.\,(\cite{cohen99}), except for $\epsilon$\,Ind, HR\,4050 and 39\,Eri\,A.
\item[$^{\mathrm{c}}$] Linear limb darkening coefficients factors from Claret et al.\,(\cite{claret95}).
\item[$^*$] The angular diameters of $\epsilon$\,Ind, HR\,4050, HR\,4546 and 39\,Eri\,A
have been measured separately with VINCI.
\end{list}
\end{table*}
%
%
% __________________ Section: Data Quality
%
\section{Data quality \label{data_quality}}
\subsection{General remarks}
Due to the fact that we used two types of light collectors (siderostats and UTs)
and several baselines (from 66 to 140\,m in ground length), the intrinsic
quality of our data is relatively heterogeneous. In this Section, we discuss briefly
the characteristics of our observations of each target. One particularity of our
measurements is that they have all been obtained during the commissioning period
of the VLTI, during which technical tasks were given higher priority.
In particular, the long baseline B3-M0 was only available during a few months
over the two years of operations of the VLTI with VINCI.
The UT1-UT3 observations were executed during two short commissioning runs
and it was not possible to obtain more than one or two phases for the observed stars
($\beta$\,Dor and $\zeta$\,Gem). However, the very large SNR values
provided by the large aperture of the UTs, even without high-order adaptive optics,
gave high-precision visibility measurements.
%______________ Figure
\begin{figure}[t]
\centering
\includegraphics[bb=0 0 360 288, width=8.5cm]{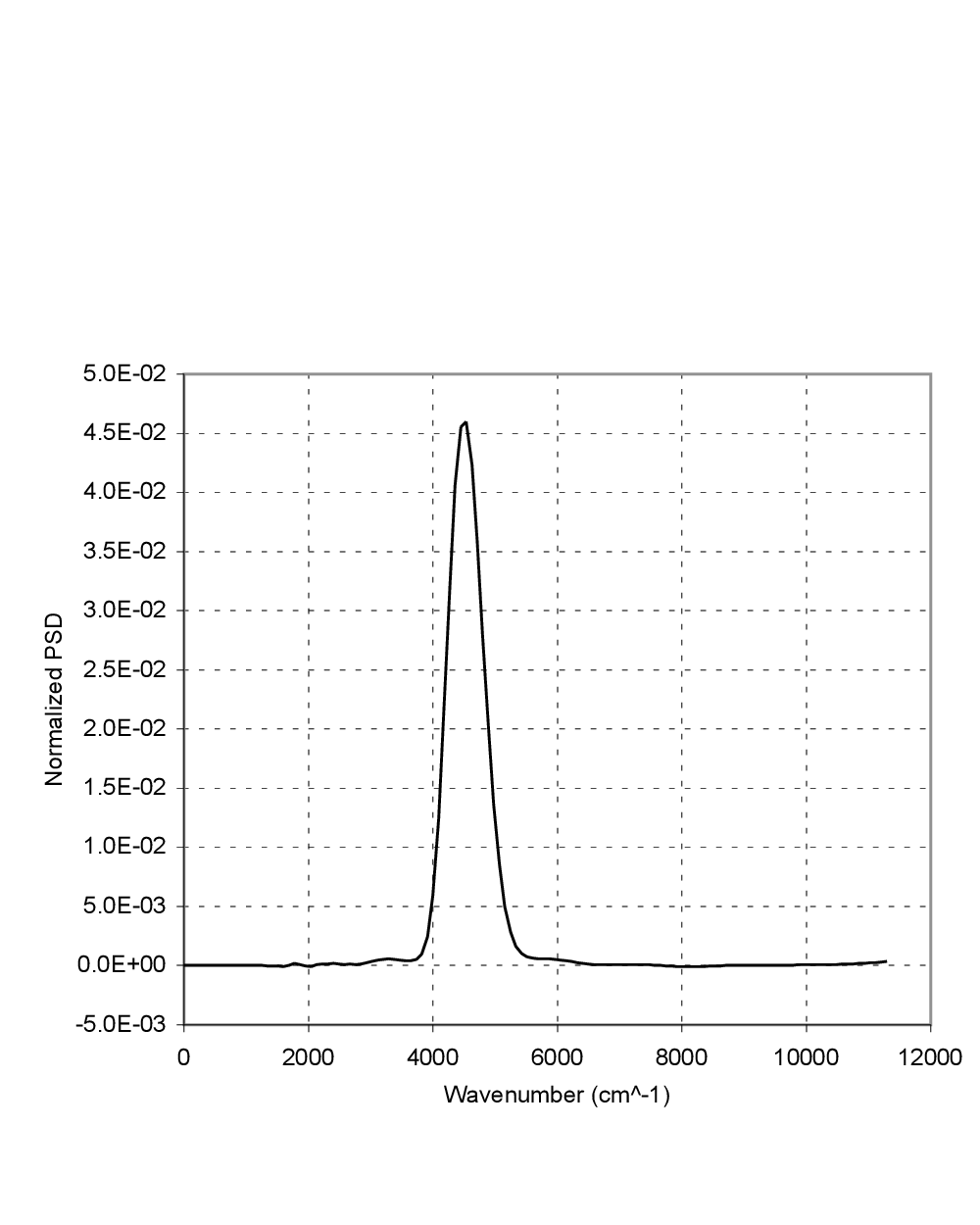}
\caption{Average wavelets power spectral density of 302 interferograms
obtained on X\,Sgr on JD=2452768.8462.
No background or bias is present. The integration of the
fringes modulated power is done between 2000 and 8000\,cm$^{-1}$}
\label{wpsd_xsgr}
\end{figure}

The VINCI processing pipeline produces a number of outputs to the user for
the data quality control, including in particular the average wavelet power spectral
density (WPSD) of the processed interferograms. This is an essential tool to verify that no
bias is present in the calibrated and normalized fringe power peak.
Fig.\,\ref{wpsd_xsgr} shows the average WPSD of a series of 302 interferograms
obtained on X\,Sgr. No bias is present, and the residual background is very low.
The power integration being done between 2000 and 8000\,cm$^{-1}$, the
complete modulated power of the fringes is taken into account without bias.

\subsection{X Sgr, W\,Sgr and Y\,Oph}
X\,Sgr was observed 8 times on the B3-M0 baseline (140\,m ground length), using
exclusively the two 0.35\,m Test Siderostats (TS).
The projected baseline length varied between
118.4 and 139.7\,m, and the observed squared visibilities were confined between
$V^2 = 56.9$ and 71.1\,\%. Thanks to its declination of $\delta = -28\,\deg$, X\,Sgr culminates
almost at zenith over Paranal (-24\,$\deg$), and all the observations were
obtained at very low airmasses. It is located on the sky near two other Cepheids of our
sample, Y\,Oph and W\,Sgr, and these three stars share the same calibrator, $\chi$\,Sco.
The average signal to noise ratio (SNR) was typically 2 to 5 on the photometric outputs of
VINCI, and 4 to 6 on the interferometric channels, for a constant fringe frequency of 242\,Hz.
A total of 4\,977 interferograms were processed by the pipeline.
The same remarks apply to W\,Sgr and Y\,Oph, as they have almost the same
magnitude and similar angular diameters. The number of processed interferograms
for these two stars was 4\,231 and 2\,182, respectively, during 9 and 4 observing sessions.

\subsection{$\eta$\,Aql}
$\eta$\,Aql was observed once on the E0-G1 baseline (66\,m) and 10 times on the
B3-M0 baseline (140\,m ground length). The total number of processed interferograms
is 5\,584. The SNRs were typically 4 and 7 on the photometric and interferometric
outputs at a fringe frequency of 242 to 272\,Hz.
Due to its northern declination ($\delta = +1\,\deg$)
and to the limits of the TS, it was not possible to observe $\eta$\,Aql for 
more than two hours per night, therefore limiting the number of interferograms and
the precision of the measurements.

\subsection{$\beta$\,Dor}
$\beta$\,Dor is a difficult target for observation with the TS, as it is partially hidden
behind the TS periscopes that are used to direct the light into the VLTI tunnels.
This causes a partial vignetting of the beams and therefore a loss in SNR.
The data from the TS are thus of intermediate quality, considering the
brightness of this star.
It is located at a declination of $-62\,\deg$, relatively close to $\ell$\,Car, and
therefore these two stars share some calibrators.  In addition to the 5 observations
with the TS, four measurements were obtained during three commissioning runs on the
UT1-UT3 baseline. A total of 8\,129 interferograms were processed, of which 5\,187 were
acquired with the 8\,m Unit Telescopes (96 minutes spread over four nights were spent on
$\beta$\,Dor using UT1 and UT3).

\subsection{$\zeta$\,Gem}
At a declination of $+20\,\deg$, $\zeta$\,Gem is not accessible to the TS due to a
mechanical limitation. This is the reason why this star was observed only on two
occasions with UT1 and UT3, for a total of 3\,857 interferograms,
obtained during 41 minutes on the target. The average on-source SNRs were typically
50 for the interferometric channels and 30 for the photometric signals, at a
fringe frequency of 694\,Hz.

The data obtained using the FLUOR/IOTA instrument are described
in Kervella et al.\,(\cite{kervella01b}). They were reprocessed using the latest release of
the FLUOR software that includes a better treatment of the photon shot noise than
the 2001 version. As the baseline of IOTA is limited to 38\,m, the visibility of the fringes
is very high, and the precision on the angular diameter is reduced compared to
the 102.5\,m baseline UT1-UT3.

\subsection{$\ell$\,Car}
As for $\beta$\,Dor, the observation of $\ell$\,Car ($\delta = -62\,\deg$) is made
particularly difficult by the vignetting of the TS beams.
Thanks to its brightness ($K \approx 1$) the SNRs are 15-20 on the interferometric
channels, and 10-15 on the photometric signals, using the TS and a fringe frequency of 242\,Hz.
One observation was obtained on the E0-G1 baseline (66\,m ground length),
and 19 measurements on the B3-M0 baseline. $\ell$\,Car is the most observed star in our
sample, with a total of 22\,226 processed interferograms. Its average diameter of approximately
3\,mas makes it an ideal target for observations with baselines of 100 to 200\,m. On the
B3-M0 baseline, we achieved projected baselines of 89.7 to 135.0\,m, corresponding
to $V^2$ values of 8 to 42\,\%. This range is ideal to constrain the visibility model
and derive precise values of the angular diameter.

\begin{figure}[t]
\centering
\includegraphics[bb=0 0 360 288, width=8.5cm]{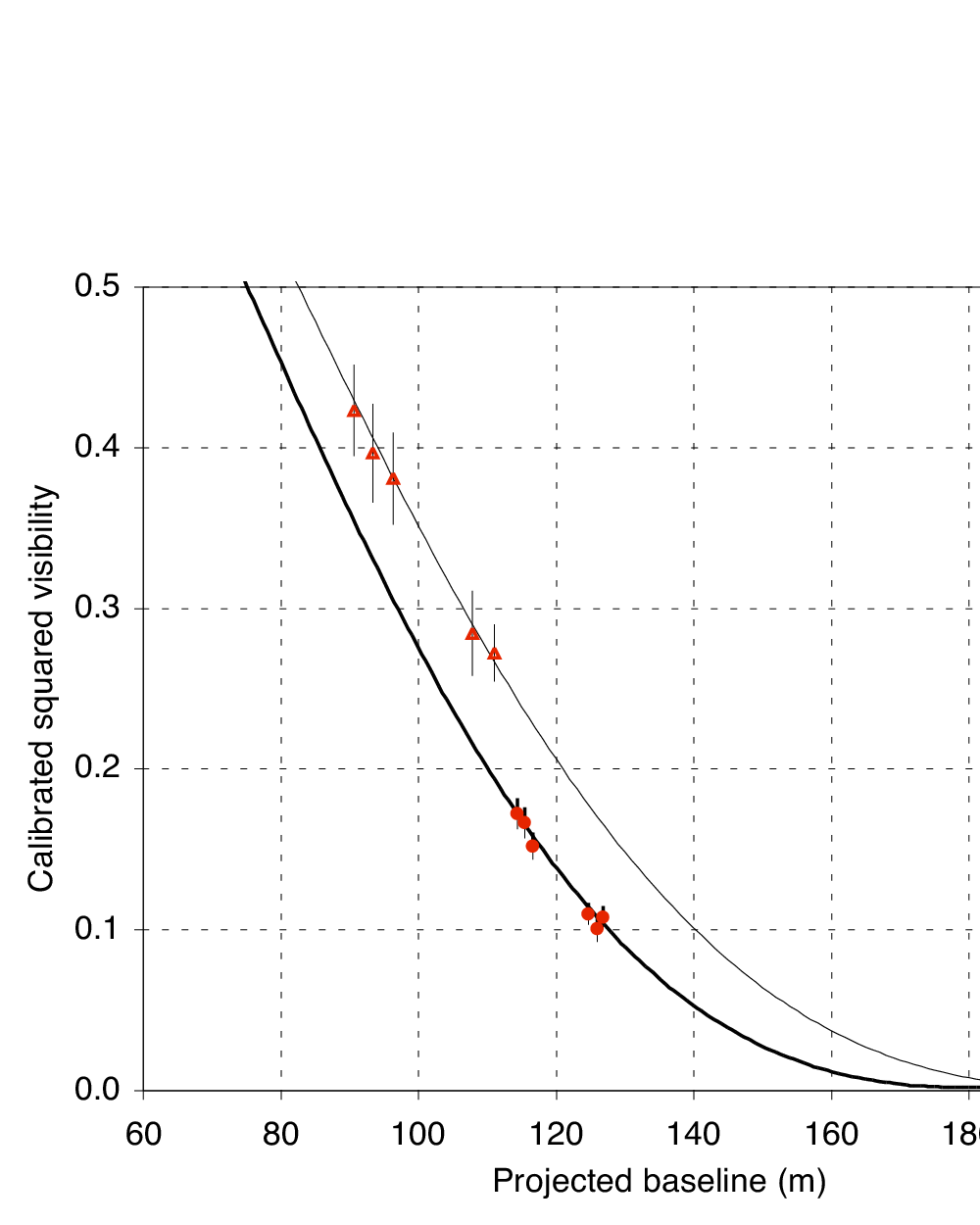}
\caption{Squared visibilities obtained on $\ell$\,Car on ${\rm JD} = 2452742.712$
(thin line) and $2452763.555$ (thick line), respectively at pulsation phases 0.722 and 0.308.
The two UD visibility models correspond to $\theta_{\rm UD} = 2.801$ and 3.075\,mas, and
take the bandwidth smearing effect into account.
The first minimum of the visibility function (that never goes down to zero)
occurs for baselines of approximately 199 and 181\,m, for an effective
wavelength of $2.18\,\mu$m.
}
\label{l_car_vis_global}
\end{figure}

Fig.~\ref{l_car_vis_global} shows the squared visibility points obtained
at two phases on $\ell$\,Car. The change in angular diameter is clearly visible.
Thanks to the variation of the projected baseline on sky, we have sampled a
segment of the visibility curve.

%______________ Figure
%
%_______________________________________________________ Section
%
\section{Angular diameters \label{angdiam}}
The object of this section is to derive the angular diameters of the Cepheids as a function
of their pulsational phase. We discuss the different types of models that can be used to
compute the angular diameter from the squared visibility measurements.
\subsection{Uniform disk angular diameters}
This very simple, rather unphysical model is commonly used for interferometric studies
as it is independent of any stellar atmosphere model. The relationship between the
visibility $V$ and the uniform disk angular diameter (UD) is:
\begin{equation}
V (B, \theta_{\rm UD}) = \left|\frac{2 {\rm J_{1}}(x)}{x}\right|
\end{equation}
where $x = \pi B\,\theta_{\rm UD} / \lambda$ is the spatial frequency. This function can
be inverted numerically to retrieve the uniform disk angular diameter $\theta_{\rm UD}$.

While the true stellar light distributions depart significantly from the UD model,
the UD angular diameters $\theta_{\rm UD}$ given in
Tables~\ref{table_angdiams_x_sgr} to \ref{table_angdiams_l_car}
have the advantage that they can easily be converted to LD values using any stellar
atmosphere model. This is achieved by computing a conversion factor
$\theta_{\rm LD} / \theta_{\rm UD}$ from the chosen intensity profile
(see e.g. Davis et al.~\cite{davis00} for details).

\subsection{Static atmosphere intensity profile \label{static_atmos}}
\begin{figure}[t]
\centering
\includegraphics[bb=0 0 360 144, width=8.5cm]{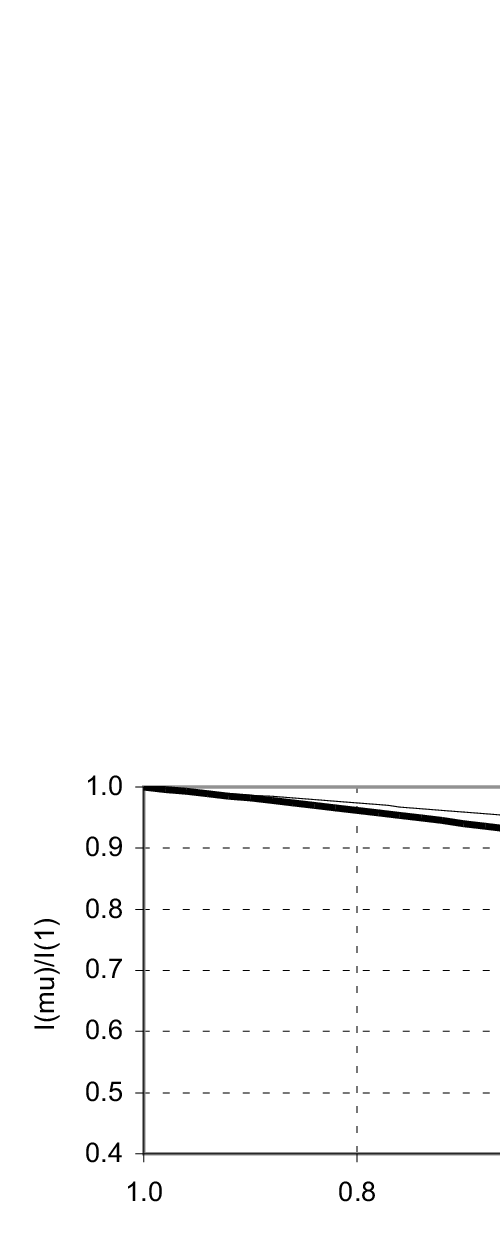}
\caption{Average intensity profiles computed from the four-parameter approximations
of Claret~(\cite{claret00}) for X\,Sgr (thin line) and $\ell$\,Car (thick line),
using the parameters listed in Table\,\ref{cepheids_params}}
\label{ld_xsgr_lcar}
\end{figure}

The visibility curve shape before the first minimum is almost impossible
to distinguish between a uniform disk (UD) and limb darkened (LD) model. Therefore, it
is necessary to use a model of the stellar disk limb darkening to deduce the photospheric angular
size of the star, from the observed visibility values.
The intensity profiles that we chose were computed by Claret\,(\cite{claret00}), based on
model atmospheres by Kurucz\,(\cite{kurucz92}).
They consist of four-parameter approximations to the function $I(\mu)/I(1)$, where
$\mu = \cos \theta$ is the cosine of the azimuth of a surface element of the star. They are
accurate approximations of the numerical results from the ATLAS modeling code.
The analytical expression of these approximations is given by:
\begin{equation}
{I(\mu)}/{I(1)} = 1 - \sum_{k=1}^{4}{a_k(1-\mu^{\frac{k}{2}}})
\end{equation}
The $a_k$ coefficients are tabulated by Claret\,(\cite{claret00}) for a wide range
of stellar parameters ($T_{\rm eff}$, $\log g$,...) and photometric bands ($U$ to $K$).
The $a_k$ values for each Cepheid are given in Table\,\ref{cepheids_params}
for the $K$ band, and the intensity profiles $I(\mu)/I(1)$ of X\,Sgr and $\ell$\,Car
are shown on Fig.\,\ref{ld_xsgr_lcar}.
%
%______________ Figure

The limb darkening is directly measurable by interferometry around the first minimum
of the visibility function, as demonstrated by several authors on giant stars
(Quirrenbach et al.~\cite{quirrenbach96}, Wittkowski et al.~\cite{wittkowski01}).
Unfortunately, even for $\ell$\,Car observed in the $K$ band, this requires a baseline of more than 180\,m
that was not available for the measurements reported here. It is intended in the near future
to measure directly the LD of a few nearby Cepheids, using the shorter wavelength bands of
AMBER (Petrov et al.\,\cite{petrov00}) and the longest baselines of the VLTI (up to 202~m).

\subsection{Changes of limb darkening with phase}
As shown by Marengo et al.\,(\cite{marengo02}),
the atmosphere of the Cepheids departs from that of a non-variable
giant with identical $T_{\rm eff}$ and $\log g$, due in particular to the
presence of energetic shock waves at certain phases of the pulsation.

However, this effect is enhanced at visible wavelengths compared to the
infrared, and appears to be negligible in the case of the VINCI observations.
Marengo et al.\,(\cite{marengo03}) have derived in the $H$ band a relative
variation of the limb darkening coefficient $k = \theta_{\rm UD} / \theta_{\rm LD}$
of only 0.2\,\%. This is below the precision of our measurements and is
neglected in the rest of this paper. Furthermore, the VINCI/VLTI measurement
wavelength being longer (2.18\,$\mu$m) than the $H$ band, the LD correction
is even smaller, as is its expected variation.

From the results of Marengo et al.\,(\cite{marengo03}) it appears clearly that
the interferometers operating at infrared wavelengths are ideally suited for
Cepheid measurements that aim at calibrating the P--R and P--L relations.
On the other hand, as pointed out by these authors, the visible wavelength
interferometers should be favored to study the dynamical evolution of the
atmosphere (including the limb darkening) during the pulsation.
The geometrical determination of the pulsation parallax is almost independant
of the adopted atmosphere model in the $K$ band, while this is not the case
at shorter wavelengths.

\subsection{Visibility model and limb darkened angular diameters \label{smearing}}
The VINCI instrument bandpass corresponds to the $K$ band filter, transparent between
$\lambda = 2.0$ and $2.4\,\mu$m.
An important effect of this relatively large spectral bandwidth
is that several spatial frequencies are simultaneously
observed by the interferometer. This effect is known as {\it bandwidth smearing}
(Kervella et al.\,\cite{kervella03b}).

To account for the bandwidth smearing, the model visibility is computed for regularly spaced
wavenumber spectral bins over the $K$ band, and then integrated to obtain the model visibility.
In this paper, we assume that the limb darkening law does not change over the $K$ band. This is
reasonable for a hot and compact stellar atmosphere, but is also coherent with the range of visibilities
measured on the Cepheids of our sample.
If necessary, this computation can easily be extended to a wavenumber dependant
$I(\mu, \sigma)$ intensity profile.
Following Davis et al.~(\cite{davis00}), using a Hankel integral, we can derive the
visibility law $V(B, \theta_{\rm LD}, \sigma)$ from the intensity profile:
\begin{equation}
V =\frac{1}{A} \
{\int_0^1{I(\mu) J_0\left({\pi B\, \sigma \, \theta_{\rm LD} \sqrt{1- \mu^2}} \right)} \mu \ d\mu}
\end{equation}
where $\sigma$ is the wavenumber:
\begin{equation}
\sigma = 1 / \lambda
\end{equation}
and $A$ is a normalization factor:
\begin{equation}
A = {\int_0^1{I(\mu) \mu \, d\mu}} 
\end{equation}

The integral of the binned squared visibilities is computed numerically
over the $K$ band and gives the model $V^2$ for the
projected baseline $B$ and the angular diameter $\theta_{\rm LD}$ through the relation:
\begin{equation}\label{eq_v2model}
V^2(\theta_{\rm LD}, B) = \int_K { \left[ V(B, \theta_{\rm LD}, \sigma)\ T(\sigma) \right] ^2 \, d\sigma }
\end{equation}
where $T(\sigma)$ is the normalized instrumental transmission defined so that
\begin{equation}
\int_K T(\sigma)\,d\sigma = 1
\end{equation}
We computed a model of $T(\sigma)$ by taking into account the instrumental
transmission of VINCI and the VLTI. It was first estimated by considering all known
factors (filter, fibers, atmospheric transmission,...) and then calibrated on sky based
on several observations of bright stars with the 8 meter UTs (see Kervella et al.~\cite{kervella03b}
for more details). This gives, for our sample of Cepheids, a measurement wavelength of
$2.179 \pm 0.003\,\mu$m. The variation of effective temperature between the stars of
our sample and over the pulsation does not change this value by more than $\pm 0.001\,\mu$m.
The uncertainty on the effective wavelength of the measurement translates to a 0.15\,\%
uncertainty on the measured angular diameters. Considering the level of the other sources
of error (statistical and systematic), the effect on our angular diameter results is negligible.

The $V^2(\theta_{\rm LD}, B)$ model is adjusted numerically to the
observed $(B, V^2)$ data using a classical $\chi^2$ minimization
process to derive $\theta_{\rm LD}$. A single angular diameter is
derived per observation session, the fit being done directly on
the set of $V^2$ values obtained during the session. The
systematic and statistical errors are considered separately in the
fitting procedure, to estimate the contribution
of the uncertainty of the calibrator diameter on the final error bar.

Each observation session was generally executed in less than 3 hours, a
short time compared to the pulsation periods of the Cepheids of our sample.
Therefore, we do not expect any phase induced smearing from this averaging.
\subsection{Measured angular diameters}
The derived angular diameters are given in Tables~\ref{table_angdiams_x_sgr}
to \ref{table_angdiams_l_car} for the seven Cepheids of our sample.
Two error bars are given for each angular diameter value:
\begin{itemize}
\item one statistical uncertainty, computed from the dispersion of the $V^2$
values obtained during the observation,
\item one systematic uncertainty defined by the error bars on
the calibrator stars {\it a priori} angular sizes.
\end{itemize}
While the statistical error can be diminished by repeatedly observing the target, the systematic
error is not reduced by averaging measurements obtained using the same calibrator.

The reference epochs $T_0$ and periods $P$ for each Cepheid are given in
Table~\ref{cepheids_params}. $N$ is the number of batches (500 interferograms)
recorded during the corresponding observing session.
For each angular diameter, the statistical and systematic calibration errors are given separately,
except for the FLUOR/IOTA measurements of $\zeta$\,Gem, for which the systematic calibration
error is negligible compared to the statistical uncertainty.
%
%
%___________________Table of angular diameter measurements
\begin{table*}
\caption{VINCI/VLTI angular diameter measurements of X\,Sgr.\label{table_angdiams_x_sgr}}
\begin{tabular}{lcccccccl}
\hline
\noalign{\smallskip}
JD & Stations & Baseline & Phase & $\theta_{\rm UD}$ (mas) & $\theta_{\rm LD}$ (mas) &
$N$ & $\chi^2_{\rm red}$ & Calibrators\\
 & & (m) & & $\pm$ stat. $\pm$ syst. & $\pm$ stat. $\pm$ syst. & & \\
\hline
\noalign{\smallskip}
%X\,Sgr -- VERSION CORRIGEE DU 01/09/03 \\
2452741.903 & B3-M0 & 138.366 & 0.560 & $1.458 \pm 0.048 \pm 0.032$ & $1.495 \pm 0.049 \pm 0.033$ & 2 & 0.66 & $\chi$\,Sco \\
2452742.885 & B3-M0 & 137.432 & 0.700 & $1.511 \pm 0.058 \pm 0.034$ & $1.549 \pm 0.059 \pm 0.035$ & 3 & 0.52 & $\chi$\,Sco \\
2452743.897 & B3-M0 & 137.903 & 0.844 & $1.415 \pm 0.055 \pm 0.034$ & $1.451 \pm 0.057 \pm 0.035$ & 3 & 0.08 & $\chi$\,Sco \\
2452744.868 & B3-M0 & 139.657 & 0.983 & $1.460 \pm 0.051 \pm 0.029$ & $1.497 \pm 0.052 \pm 0.030$ & 2 & 0.09 & $\chi$\,Sco \\
2452747.848 & B3-M0 & 139.530 & 0.408 & $1.499 \pm 0.213 \pm 0.038$ & $1.537 \pm 0.219 \pm 0.039$ & 1 & - & $\chi$\,Sco \\
2452749.832 & B3-M0 & 139.084 & 0.691 & $1.429 \pm 0.099 \pm 0.034$ & $1.465 \pm 0.101 \pm 0.034$ & 2 & 0.35 & $\chi$\,Sco \\
2452766.811 & B3-M0 & 138.853 & 0.112 & $1.393 \pm 0.070 \pm 0.036$ & $1.428 \pm 0.071 \pm 0.037$ & 4 & 0.09 & $\chi$\,Sco \\
2452768.877 & B3-M0 & 128.228 & 0.406 & $1.413 \pm 0.016 \pm 0.028$ & $1.449 \pm 0.016 \pm 0.029$ & 6 & 0.62 & $\chi$\,Sco \\
\noalign{\smallskip}
\hline
\end{tabular}
\end{table*}
%
%
%%___________________Table of angular diameter measurements
\begin{table*}
\caption{Angular diameter measurements of $\eta$\,Aql.\label{table_angdiams_eta_aql}}
\begin{tabular}{lcccccccl}
\hline
\noalign{\smallskip}
JD & Stations & Baseline & Phase & $\theta_{\rm UD}$ (mas) & $\theta_{\rm LD}$ (mas) &
$N$ & $\chi^2_{\rm red}$ & Calibrators\\
 & & (m) & & $\pm$ stat. $\pm$ syst. & $\pm$ stat. $\pm$ syst. & & \\
\hline
\noalign{\smallskip}
%$\eta$\,Aql -- VERSION CORRIGEE DU 01/09/03\\
2452524.564 & E0-G1 & 60.664 & 0.741 & $1.746 \pm 0.100 \pm 0.074$ & $1.792 \pm 0.103 \pm 0.076$ & 3 & 0.08 & 70\,Aql\\
2452557.546 & B3-M0 & 137.625 & 0.336 & $1.877 \pm 0.098 \pm 0.037$ & $1.931 \pm 0.101 \pm 0.038$ & 1 & - & $\epsilon$\,Ind\\
2452559.535 & B3-M0 & 138.353 & 0.614 & $1.806 \pm 0.037 \pm 0.027$ & $1.857 \pm 0.038 \pm 0.027$ & 1 & - & 7\,Aqr, $\epsilon$\,Ind\\
2452564.532 & B3-M0 & 136.839 & 0.310 & $1.809 \pm 0.043 \pm 0.031$ & $1.860 \pm 0.045 \pm 0.032$ & 3 & 0.42 & 7\,Aqr, $\epsilon$\,Ind\\
2452565.516 & B3-M0 & 138.495 & 0.447 & $1.871 \pm 0.017 \pm 0.027$ & $1.924 \pm 0.017 \pm 0.028$ & 3 & 0.13 & 7\,Aqr\\
2452566.519 & B3-M0 & 137.845 & 0.587 & $1.861 \pm 0.023 \pm 0.026$ & $1.914 \pm 0.024 \pm 0.026$ & 5 & 0.23 & 7\,Aqr\\
2452567.523 & B3-M0 & 137.011 & 0.727 & $1.802 \pm 0.027 \pm 0.030$ & $1.853 \pm 0.028 \pm 0.030$ & 2 & 0.62 & 7\,Aqr\\
2452573.511 & B3-M0 & 136.303 & 0.561 & $1.884 \pm 0.053 \pm 0.022$ & $1.938 \pm 0.054 \pm 0.022$ & 1 & - & $\lambda$\,Gru, HR\,8685\\
2452769.937 & B3-M0 & 139.632 & 0.931 & $1.647 \pm 0.026 \pm 0.018$ & $1.693 \pm 0.026 \pm 0.018$ & 3 & 0.06 & $\chi$\,Sco \\
2452770.922 & B3-M0 & 139.400 & 0.068 & $1.791 \pm 0.041 \pm 0.027$ & $1.842 \pm 0.042 \pm 0.028$ & 3 & 0.15 & $\chi$\,Sco\\
2452772.899 & B3-M0 & 138.188 & 0.343 & $1.880 \pm 0.044 \pm 0.026$ & $1.934 \pm 0.046 \pm 0.027$ & 3 & 0.16 & 7\,Aqr \\
\noalign{\smallskip}
\hline
\end{tabular}
\end{table*}
%
%
%%___________________Table of angular diameter measurements
\begin{table*}
\caption{Angular diameter measurements of W\,Sgr.\label{table_angdiams_w_sgr}}
\begin{tabular}{lcccccccl}
\hline
\noalign{\smallskip}
JD & Stations & Baseline & Phase & $\theta_{\rm UD}$ (mas) & $\theta_{\rm LD}$ (mas) &
$N$ & $\chi^2_{\rm red}$ & Calibrators\\
 & & (m) & & $\pm$ stat. $\pm$ syst. & $\pm$ stat. $\pm$ syst. & & \\
\hline
\noalign{\smallskip}
%W\,Sgr  -- VERSION CORRIGEE DU 01/09/03\\
2452743.837 & B3-M0 & 137.574 & 0.571 & $1.408 \pm 0.096 \pm 0.038$ & $1.447 \pm 0.099 \pm 0.039$ & 1 & - & $\chi$\,Sco \\
2452744.915 & B3-M0 & 137.166 & 0.713 & $1.292 \pm 0.088 \pm 0.034$ & $1.327 \pm 0.090 \pm 0.035$ & 2 & 0.04 & $\chi$\,Sco \\
2452749.868 & B3-M0 & 139.632 & 0.365 & $1.262 \pm 0.141 \pm 0.040$ & $1.297 \pm 0.145 \pm 0.041$ & 1 & - & $\chi$\,Sco \\
2452751.866 & B3-M0 & 139.538 & 0.628 & $1.320 \pm 0.174 \pm 0.041$ & $1.357 \pm 0.179 \pm 0.042$ & 1 & - & $\chi$\,Sco \\
2452763.888 & B3-M0 & 131.830 & 0.211 & $1.284 \pm 0.019 \pm 0.029$ & $1.319 \pm 0.020 \pm 0.030$ & 4 & 0.73 & $\chi$\,Sco \\
2452764.856 & B3-M0 & 135.926 & 0.339 & $1.355 \pm 0.021 \pm 0.021$ & $1.393 \pm 0.021 \pm 0.022$ & 4 & 0.76 & $\chi$\,Sco \\
2452765.880 & B3-M0 & 132.679 & 0.473 & $1.313 \pm 0.022 \pm 0.025$ & $1.349 \pm 0.023 \pm 0.026$ & 4 & 1.43 & $\chi$\,Sco \\
2452767.867 & B3-M0 & 132.637 & 0.735 & $1.208 \pm 0.073 \pm 0.039$ & $1.241 \pm 0.075 \pm 0.040$ & 3 & 0.01 & $\chi$\,Sco \\
2452769.914 & B3-M0 & 120.648 & 0.005 & $1.240 \pm 0.055 \pm 0.034$ & $1.274 \pm 0.056 \pm 0.035$ & 2 & 0.33 & $\chi$\,Sco \\
\noalign{\smallskip}
\hline
\end{tabular}
\end{table*}
%
%
%%___________________Table of angular diameter measurements
\begin{table*}
\caption{Angular diameter measurements of $\beta$\,Dor.\label{table_angdiams_beta_dor}}
\begin{tabular}{lcccccccl}
\hline
\noalign{\smallskip}
JD & Stations & Baseline & Phase & $\theta_{\rm UD}$ (mas) & $\theta_{\rm LD}$ (mas) &
$N$ & $\chi^2_{\rm red}$ & Calibrators\\
 & & (m) & & $\pm$ stat. $\pm$ syst. & $\pm$ stat. $\pm$ syst. & & \\
\hline
\noalign{\smallskip}
%$\beta$\,Dor  -- VERSION CORRIGEE DU 01/09/03\\
2452215.795 & U1-U3 & 89.058 & 0.161 & $1.842 \pm 0.036 \pm 0.074$ & $1.896 \pm 0.036 \pm 0.074$ & 3 & 0.03 & $\chi$\,Phe, $\gamma^2$\,Vol\\
2452216.785 & U1-U3 & 89.651 & 0.261 & $1.954 \pm 0.026 \pm 0.040$ & $2.011 \pm 0.026 \pm 0.040$ & 7 & 0.10 & $\gamma^2$\,Vol\\
2452247.761 & U1-U3 & 83.409 & 0.408 & $1.921 \pm 0.045 \pm 0.039$ & $1.977 \pm 0.045 \pm 0.039$ & 5 & 0.40 & $\epsilon$\,Ret\\
2452308.645 & U1-U3 & 75.902 & 0.594 & $1.844 \pm 0.027 \pm 0.071$ & $1.897 \pm 0.027 \pm 0.071$ & 5 & 1.01 & HD\,63697\\
2452567.827 & B3-M0 & 134.203 & 0.927 & $1.793 \pm 0.039 \pm 0.049$ & $1.848 \pm 0.039 \pm 0.049$ & 1 & - & HR\,2549\\
2452744.564 & B3-M0 & 89.028 & 0.884 & $1.730 \pm 0.064 \pm 0.032$ & $1.780 \pm 0.064 \pm 0.032$ & 2 & 0.09 & HR\,3046, 4831\\
2452749.514 & B3-M0 & 98.176 & 0.387 & $1.921 \pm 0.106 \pm 0.029$ & $1.978 \pm 0.106 \pm 0.029$ & 3 & 0.11 & HR\,3046\\
2452750.511 & B3-M0 & 98.189 & 0.488 & $1.864 \pm 0.065 \pm 0.039$ & $1.919 \pm 0.065 \pm 0.039$ & 2 & 0.24 & HR\,3046\\
2452751.519 & B3-M0 & 95.579 & 0.591 & $1.954 \pm 0.169 \pm 0.030$ & $2.012 \pm 0.169 \pm 0.030$ & 3 & 0.03 & HR\,3046\\
\noalign{\smallskip}
\hline
\end{tabular}
\end{table*}
%%___________________Table of angular diameter measurements
\begin{table*}
\caption{VINCI/VLTI  and FLUOR/IOTA angular diameter measurements of $\zeta$\,Gem.
No systematic calibration error is given for FLUOR/IOTA values (negligible compared to the
statistical uncertainty). The baseline is given for the VINCI/VLTI observations (in m), while the spatial frequency (in {\it italic})
is listed for the measurements obtained with FLUOR, expressed in cycles/arcsec. \label{table_angdiams_zeta_gem}}
\begin{tabular}{lcccccccl}
\hline
\noalign{\smallskip}
JD & Stations & B, {\it SF} & Phase & $\theta_{\rm UD}$ (mas) & $\theta_{\rm LD}$ (mas) &
$N$ & $\chi^2_{\rm red}$ & Calibrators\\
 & & & & $\pm$ stat. $\pm$ syst. & $\pm$ stat. $\pm$ syst. & & \\
\hline
\noalign{\smallskip}
%$\zeta$\,Gem -- VERSION CORRIGEE DU 01/09/03\\
2452214.879 & U1-U3 & 82.423 & 0.408 & $1.677 \pm 0.030 \pm 0.051$ & $1.725 \pm 0.031 \pm 0.052$ & 8 & 0.25 & 39\,Eri\\
2452216.836 & U1-U3 & 72.837 & 0.600 & $1.712 \pm 0.057 \pm 0.067$ & $1.760 \pm 0.058 \pm 0.069$ & 6 & 0.28 & 39\,Eri, $\gamma^2$\,Vol\\
2451527.972 & IOTA-38m & {\it 84.870} & 0.739 & $1.606 \pm 0.334$ & $1.651 \pm 0.343$ & 1 & - & HD\,49968\\
2451601.828 & IOTA-38m & {\it 83.917} & 0.014 & $1.709 \pm 0.086$ & $1.795 \pm 0.088$ & 3 & 0.02 & HD\,49968\\
2451259.779 & IOTA-38m & {\it 83.760} & 0.318 & $2.040 \pm 0.291$ & $2.144 \pm 0.299$ & 1 & -  & HD\,49968\\
2451262.740 & IOTA-38m & {\it 84.015} & 0.610 & $1.692 \pm 0.273$ & $1.767 \pm 0.281$ & 2 & 0.13 & HD\,49968\\
2451595.863 & IOTA-38m & {\it 83.790} & 0.427 & $1.391 \pm 0.284$ & $1.306 \pm 0.292$ & 2 & 1.72 & HD\,49968\\
2451602.764 & IOTA-38m & {\it 85.010} & 0.107 & $1.867 \pm 0.216$ & $1.962 \pm 0.222$ & 2 & 0.02 & HD\,49968\\
\noalign{\smallskip}
\hline
\end{tabular}
\end{table*}
%%___________________Table of angular diameter measurements
\begin{table*}
\caption{Angular diameter measurements of Y\,Oph.\label{table_angdiams_y_oph}}
\begin{tabular}{lcccccccl}
\hline
\noalign{\smallskip}
JD & Stations & Baseline & Phase & $\theta_{\rm UD}$ (mas) & $\theta_{\rm LD}$ (mas) &
$N$ & $\chi^2_{\rm red}$ & Calibrators\\
 & & (m) & & $\pm$ stat. $\pm$ syst. & $\pm$ stat. $\pm$ syst. & & \\
\hline
\noalign{\smallskip}
%Y\,Oph -- VERSION CORRIGEE DU 22/11/03\\
2452742.906 & B3-M0 & 139.569 & 0.601 & $1.427 \pm 0.115 \pm 0.034$ & $1.472 \pm 0.119 \pm 0.035$ & 2 & 0.10 & $\chi$\,Sco \\
2452750.884 & B3-M0 & 139.057 & 0.067 & $1.380 \pm 0.100 \pm 0.034$ & $1.423 \pm 0.103 \pm 0.035$ & 2 & 0.41 & $\chi$\,Sco \\
2452772.831 & B3-M0 & 139.657 & 0.349 & $1.443 \pm 0.051 \pm 0.025$ & $1.488 \pm 0.053 \pm 0.026$ & 3 & 0.22 & $\chi$\,Sco \\
2452782.186 & B3-M0 & 129.518 & 0.168 & $1.402 \pm 0.027 \pm 0.037$ & $1.445 \pm 0.028 \pm 0.038$ & 4 & 0.30 & $\chi$\,Sco \\
\noalign{\smallskip}
\hline
\end{tabular}
\end{table*}
%%___________________Table of angular diameter measurements
\begin{table*}
\caption{Angular diameter measurements of $\ell$\,Car.\label{table_angdiams_l_car}}
\begin{tabular}{lcccccccl}
\hline
\noalign{\smallskip}
JD & Stations & Baseline & Phase & $\theta_{\rm UD}$ (mas) & $\theta_{\rm LD}$ (mas) &
$N$ & $\chi^2_{\rm red}$ & Calibrators\\
 & & (m) & & $\pm$ stat. $\pm$ syst. & $\pm$ stat. $\pm$ syst. & & & HR \\
\hline
\noalign{\smallskip}
%$\ell$\,Car -- VERSION CORRIGEE DU 01/09/03\\
2452453.498 & E0-G1 & 61.069 & 0.587 & $2.958 \pm 0.039 \pm 0.102$ & $3.054 \pm 0.041 \pm 0.105$ & 4 & 0.01 & 4050 \\
2452739.564 & B3-M0 & 130.468 & 0.634 & $2.786 \pm 0.073 \pm 0.042$ & $2.891 \pm 0.076 \pm 0.043$ & 2 & 0.03 & 4526 \\
2452740.569 & B3-M0 & 128.821 & 0.662 & $2.879 \pm 0.017 \pm 0.042$ & $2.989 \pm 0.018 \pm 0.044$ & 7 & 0.77 & 4526 \\
2452741.717 & B3-M0 & 96.477 & 0.694 & $2.893 \pm 0.025 \pm 0.028$ & $2.993 \pm 0.026 \pm 0.029$ & 5 & 0.28 & 4526 \\
2452742.712 & B3-M0 & 99.848 & 0.722 & $2.801 \pm 0.034 \pm 0.042$ & $2.899 \pm 0.035 \pm 0.043$ & 5 & 0.09 & 4526 \\
2452743.698 & B3-M0 & 99.755 & 0.750 & $2.667 \pm 0.071 \pm 0.015$ & $2.758 \pm 0.074 \pm 0.016$ & 2 & 0.08 & 4831 \\
2452744.634 & B3-M0 & 114.981 & 0.776 & $2.698 \pm 0.031 \pm 0.012$ & $2.794 \pm 0.032 \pm 0.013$ & 6 & 0.73 & 4831 \\
2452745.629 & B3-M0 & 115.791 & 0.804 & $2.584 \pm 0.094 \pm 0.017$ & $2.675 \pm 0.097 \pm 0.017$ & 2 & 0.01 & 3046, 4546, 4831\\
2452746.620 & B3-M0 & 116.828 & 0.832 & $2.679 \pm 0.023 \pm 0.039$ & $2.775 \pm 0.023 \pm 0.040$ & 5 & 0.65 & 3046, 4546 \\
2452747.599 & B3-M0 & 120.812 & 0.860 & $2.606 \pm 0.122 \pm 0.025$ & $2.699 \pm 0.127 \pm 0.026$ & 3 & 0.70 & 4546, 4831 \\
2452749.576 & B3-M0 & 124.046 & 0.915 & $2.553 \pm 0.075 \pm 0.011$ & $2.645 \pm 0.077 \pm 0.012$ & 4 & 1.18 & 4546\\
2452751.579 & B3-M0 & 122.555 & 0.971 & $2.657 \pm 0.027 \pm 0.017$ & $2.753 \pm 0.028 \pm 0.017$ & 4 & 1.16 & 3046, 4831 \\
2452755.617 & B3-M0 & 112.185 & 0.085 & $2.867 \pm 0.109 \pm 0.013$ & $2.970 \pm 0.113 \pm 0.013$ & 1 & - & 4831 \\
2452763.555 & B3-M0 & 120.632 & 0.308 & $3.077 \pm 0.008 \pm 0.031$ & $3.194 \pm 0.009 \pm 0.033$ & 6 & 1.02 & 4546 \\
2452765.555 & B3-M0 & 119.629 & 0.365 & $3.094 \pm 0.011 \pm 0.031$ & $3.212 \pm 0.011 \pm 0.033$ & 6 & 1.19 & 4546 \\
2452766.550 & B3-M0 & 120.005 & 0.393 & $3.092 \pm 0.011 \pm 0.032$ & $3.210 \pm 0.011 \pm 0.033$ & 7 & 0.99 & 4546 \\
2452768.566 & B3-M0 & 115.135 & 0.450 & $3.075 \pm 0.010 \pm 0.034$ & $3.188 \pm 0.011 \pm 0.035$ & 7 & 0.46 & 4546 \\
2452769.575 & B3-M0 & 113.082 & 0.478 & $3.075 \pm 0.018 \pm 0.011$ & $3.189 \pm 0.018 \pm 0.012$ & 3 & 0.03 & 3046, 4831 \\
2452770.535 & B3-M0 & 121.152 & 0.505 & $3.044 \pm 0.019 \pm 0.009$ & $3.160 \pm 0.020 \pm 0.009$ & 2 & 0.20 & 3046, 4831 \\
2452771.528 & B3-M0 & 122.014 & 0.533 & $3.021 \pm 0.017 \pm 0.010$ & $3.136 \pm 0.017 \pm 0.010$ & 3 & 0.88 & 4831\\
\noalign{\smallskip}
\hline
\end{tabular}
\end{table*}
%
%_______________________________________________________ Section
%
\section{Linear diameter curves \label{radius_curves_sect}}
For each star we used radial velocity data found in the literature.
Specifically, we collected data from Bersier\,(\cite{bersier02}) for $\eta$\,Aql,
$\ell$\,Car, and $\beta$\,Dor; from Bersier et al.\,(\cite{bersier94}) for $\zeta$\,Gem;
from Babel et al\,(\cite{babel89}) for W\,Sgr. All these data have been
obtained with the CORAVEL radial velocity spectrograph
(Baranne, Mayor \& Poncet\,\cite{baranne79}).
We also obtained data from Evans \& Lyons\,(\cite{evans86}) for Y\,Oph
and from Wilson et al.\,(\cite{wilson89}) for X\,Sgr.

In theory, the linear diameter variation could be determined by direct integration of
pulsational velocities
(within the assumption that the $\tau = 1$ photosphere is comoving
with the atmosphere of the Cepheid during its pulsation).
However these velocities are deduced from the measured radial
velocities by the use of a projection factor $p$. The Cepheid's radii determined
from the BW method depend directly from a good knowledge of $p$.
Sabbey et al.\,(\cite{sabbey95}) and
Krockenberger et al.\,(\cite{krockenberger97}) have studied
in detail the way to determine the $p$-factor.
We used a constant projection factor $p = 1.36$ in order to transform the radial
velocities into pulsation velocities. Burki et al.\,(\cite{burki82}) have
shown that this value is appropriate for the radial velocity measurements that
we used.

%___________________________________________________ Section
\section{Cepheids parameters\label{puls_param}}
\subsection{Angular diameter model fitting and distance measurement}
From our angular diameter measurements, we can
derive both the average linear diameter and the distance to the Cepheids.
This is done by applying a classical $ {\chi}^{2} $ minimization algorithm 
between our angular diameter measurements and a model of the star
pulsation. The minimized quantity with respect to the chosen model is
\begin{equation}
\label{chi2sum}
{\chi}^{2} = \sum_{i}{\frac{(\theta_{\rm LD\,observ}(\phi_{i}) -
\theta_{\rm LD\,model}(\phi_{i}))^2}{\sigma_{\rm observ}(\phi_{i})^2}}
\end{equation}
where $\phi_{i}$ is the phase of measurement $i$.
The expression of $\theta_{\rm LD\,model}(\phi_{i})$ is defined using the
following parameters: 
\begin{itemize}
\item the average LD angular diameter $\overline{\theta_{\rm LD}}$ (in mas),
\item the linear diameter variation $\Delta D(\phi_{i})$ (in $D_{\odot}$),
\item the distance $d$ to the star (in pc).
\end{itemize}
The resulting expression is therefore:
\begin{equation}\label{diam_model_eq}
\theta_{\rm LD\,model}(\phi_{i}) = \overline{\theta_{\rm LD}} + 9.305\,\left(\frac{\Delta D(\phi_{i})}{d}\right) [{\rm mas}]
\end{equation}
As $\Delta D(\phi_{i})$ is known from the integration of the radial velocity
curve (Sect.\,\ref{radius_curves_sect}), the only variable parameters are the
average LD angular diameter $\overline{\theta_{\rm LD}}$ and the distance $d$.
From there, three methods can be used to derive the distance $d$, depending
on the level of completeness and precision of the angular diameter measurements:
\begin{itemize}
\item{{\bf Constant diameter fit (order 0):}
The average linear diameter $\overline{D}$ of the star is supposed known {\it a priori}
from previously published BW measurements or P--R relations (See
Sect.~\ref{published_diams}). We assume here that $\Delta D (\phi) = 0$.
The only remaining variable to fit is the distance $d$.
This is the most basic method, and is useful as a
reference to assess the level of detection of the pulsational
diameter variation with the other methods.}
\item{{\bf  Variable diameter (order 1):}
We still consider that the average linear diameter $\overline{D}$
of the star is known {\it a priori}, but we include in our model the
radius variation derived from the integration of the radial velocity curve.
This method is well suited when the intrinsic accuracy of the angular
diameter measurements is too low to measure precisely the pulsation
amplitude ($\zeta$\,Gem, X\,Sgr and Y\,Oph).
The distance $d$ is the only free parameter for the fit.}
\item{{\bf Complete fit (order 2):}
The average LD angular diameter $\overline{\theta_{\rm LD}}$
and the distance $d$ are both considered as variables and adjusted
simultaneously to the angular diameter measurements.
In the fitting process, the radius curve is matched to the observed
pulsation amplitude.
Apart from direct trigonometric parallax, this implementation of the BW
method is the most direct way of measuring the distance
and diameter of a Cepheid.
It requires a high precision angular diameter curve and a
good phase coverage.
It can be applied directly to our $\eta$\,Aql, W\,Sgr, $\beta$\,Dor and
$\ell$\,Car measurements.}
\end{itemize}
\subsection{Published linear diameter values \label{published_diams}}
In this section, we survey the existing linear diameter determinations
for the Cepheids of our sample, in order to apply the order 0 and 1 methods
to our observations.

A large number of BW studies have been published, using both visible and
infrared wavelength observations. For $\zeta$\,Gem and $\eta$\,Aql, the pulsation
has been resolved using the Palomar Testbed Interferometer (Lane et al.\,\cite{lane00},
\cite{lane02}), thererefore giving a direct estimate of the diameter and distance of these
stars. Table\,\ref{published_cepheids} gives a list of the existing diameter
estimates for the Cepheids of our sample from the application of the classical
BW method (``{\sc Baade-Wesselink}'' section of the table).

From the many different P--R relations available, we chose the
Gieren et al.\,(\cite{gieren98}) version, as it is based on infrared colors for
the determination of the temperature of the stars. Compared to visible colors,
the infrared colors give a much less dispersed P--R relation.
Indeed, this relation has a very good intrinsic precision of the order of 5 to 10\,\%
for the period range of our sample. Moreover, it is identical
to the law determined by Laney \& Stobie\,(\cite{laney95}).
The compatibility  with the individual BW diameter estimates is also satisfactory.
The linear diameters deduced from this P--R law are mentioned in
the``{\sc Empirical P--R}" section of Table\,\ref{published_cepheids}.
We assume these linear diameter values in the following.
%
%___________________Table of published diameters
\begin{table*}
\caption{Published linear diameter estimates, expressed in $D_{\rm \odot}$.}
\label{published_cepheids}
\begin{tabular}{llllllll}
\hline
 & X~Sgr & $\eta$~Aql & W~Sgr & $\beta$~Dor & $\zeta$~Gem & Y~Oph & $\ell$~Car \\
\hline
{\sc Interferometry} \\
Kervella et al.\,(\cite{kervella01b})$^{\mathrm{*}}$ & & & & & $63^{+35}_{-19}$ & & \\
Lane et al.\,(\cite{lane02}) & & $61.8 \pm 7.6$ & & & $66.7 \pm 7.2$ & & \\
Nordgren et al.\,(\cite{nordgren00})$^{\mathrm{*}}$ & & $69^{+28}_{-15}$ & & & $60^{+25}_{-14}$ & & \\
\noalign{\smallskip}
%\hline
%{\sc Mean Interf.} &  & 62.7 &  &  & 65.8 &  &  \\
\hline \hline
\noalign{\smallskip}
{\sc Baade-Wesselink} \\
Bersier et al.\,(\cite{bersier97}) & & & $56.0 \pm 2.9$ & & $89.5 \pm 13.3$ & & \\
Fouqu\'e et al.\,(\cite{fouque03}) & & $48.1 \pm 1.1$ & & & & & $201.7 \pm 3.0$ \\
Krockenberger et al.\,(\cite{krockenberger97}) & & & $56.8 \pm 2.3$ & & $69.1^{+5.5}_{-4.8}$ & & \\
Laney\,\&\,Stobie\,(\cite{laney95}) & & & & $63.5 \pm 1.8$ & & $92.2 \pm 3.2$ & $180.1 \pm 4.5$ \\
Moffett\,\&\,Barnes\,(\cite{moffett87})$^{\mathrm{a}}$ & $47.8 \pm 4.5$ & $52.8 \pm 3.8$ &
$60.8 \pm 7.6$ & & $62.6 \pm 11.5$ & & \\
Moffett\,\&\,Barnes\,(\cite{moffett87})$^{\mathrm{b}}$ & $49.6 \pm 4.6$ & $54.8 \pm 3.9$ &
$63.1 \pm 7.8$ & & $64.9 \pm 11.9$ & & \\
Sabbey et al.\,(\cite{sabbey95})$^{\mathrm{c}}$ & $42.2 \pm 4.1$ & $62.7 \pm 3.1$ & & & $61.8 \pm 3.5$ & & \\
Sabbey et al.\,(\cite{sabbey95})$^{\mathrm{d}}$ & $66.6 \pm 4.9$ & $65.8 \pm 3.2$ & & & $64.4 \pm 3.6$ & & \\
Sachkov et al.\,(\cite{sachkov98}) & & & & & $74 \pm 10$ & & \\
Taylor et al.\,(\cite{taylor97}) & & & & & & & $179.2 \pm 10.4$\\
Taylor\,\&\,Booth\,(\cite{taylor98}) & & & & $67.8 \pm 0.7$ & & & \\
Turner\,\&\,Burke\,(\cite{turner02}) & & $52.6 \pm 8.9$ & & $53.8 \pm 1.9$ & & & \\
Sasselov\,\&\,Lester\,(\cite{sasselov90}) & $67 \pm 6$ & $62 \pm 6$ & & & & & \\
\noalign{\smallskip}
\hline
\noalign{\smallskip}
{\sc Mean B--W} (overall $\sigma$) & 52.5 (11.4) & 59.9 (5.7) & 57.0 (3.4) & 65.8 (7.2) & 65.3 (9.8) & 92.2 (-) & 180 (-) \\
\hline \hline
\noalign{\smallskip}
{\sc Empirical P--R} \\
Gieren et al.\,(\cite{gieren98}) & $ 51.2 \pm 2.6$ & $ 52.1 \pm 2.7$ &
$ 54.4 \pm 2.9$ & $ 66.0 \pm 3.9$ & $ 67.6 \pm 4.0$ & $ 100.1 \pm 7.3$ & $ 173.1 \pm 15.8$ \\
\noalign{\smallskip}
\hline \hline
\end{tabular}
\begin{list}{}{}
\item[$^{\mathrm{*}}$] $\zeta$\,Gem values were derived from Kervella et al.\,(\cite{kervella01b})
and Nordgren et al.\,(\cite{nordgren00}) using the {\sc Hipparcos} parallaxes.
$\eta$\,Aql was taken from Nordgren et al.\,(\cite{nordgren00})
\item[$^{\mathrm{a}}$] Assuming a constant $p$--factor.
\item[$^{\mathrm{b}}$] Assuming a variable $p$--factor.
\item[$^{\mathrm{c}}$] Bisector method.
\item[$^{\mathrm{d}}$] Parabolic fit method.
\end{list}
\end{table*}
\subsection{Angular diameter fitting results}
The results of both constant and variable diameter fits for the seven Cepheids
of our sample are listed in Table\,\ref{results_cepheids0} to \ref{results_cepheids2}.
$\eta$\,Aql, W\,Sgr, $\beta$\,Dor and $\ell$\,Car gave results for all fitting methods,
while X\,Sgr, $\zeta$\,Gem and Y\,Oph were limited to order 1 models.
For X\,Sgr, the order 1 fit is less adequate than the order 0,
considering the quality of our measurements of this star. This is shown by the
fact that the $\chi^2$ is significantly higher for the order 1 fit (1.36) than for the
order 0 (0.38).

In the case of $\ell$\,Car, the fit of a constant diameter results in a very high
$\chi^2$ value. This means that the average diameters $\overline{\theta_{\rm UD 0}}$
and $\overline{\theta_{\rm LD 0}}$ should not be used for further analysis.
The pulsation curve of this star is not sampled uniformly by our interferometric
observations, with more values around the maximum diameter.
This causes the larger diameter values to have more weight in the average diameter
computation, and this produces a significant positive bias. This remark does not
apply to the orders 1 and 2 fitting methods.

As a remark, no significant phase shift is detected at a level of $2.5\ 10^{-4}$
(14 minutes of time) between the predicted radius curve of $\ell$\,Car
and the observed angular diameter curve. The values of $P$ and $T_0$ used for
the fit are given in Table \ref{sample_cephs}.

\begin{figure}[t]
\centering
\includegraphics[bb=0 0 360 288, width=8.5cm]{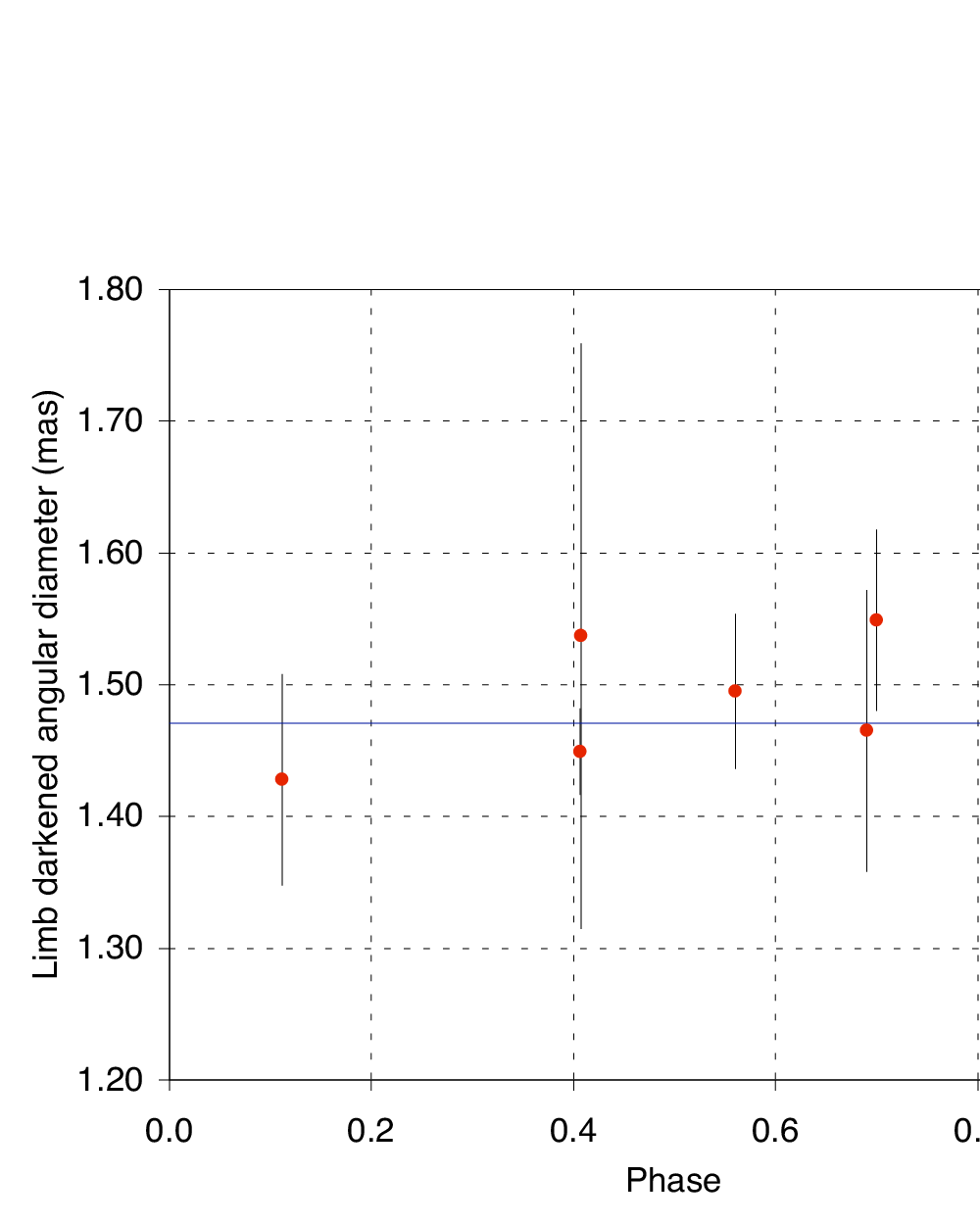}
\caption{Order 0 model fit for X\,Sgr.}
\label{x_sgr_fit}
\end{figure}

\begin{figure}[t]
\centering
\includegraphics[bb=0 0 360 288, width=8.5cm]{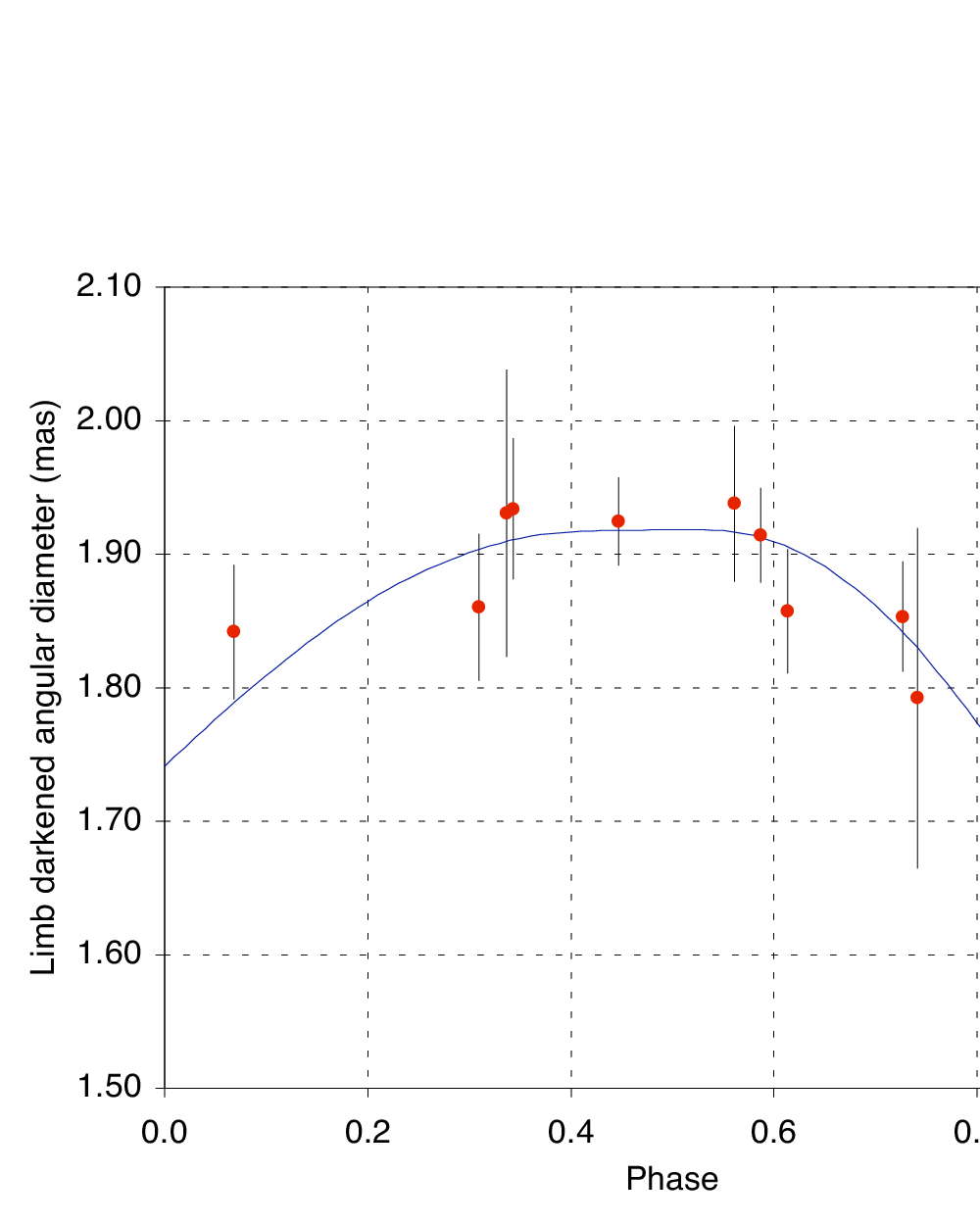}
\caption{Order 2 model fit for $\eta$\,Aql. The superimposed angular
diameter variation curve (thin line) is derived from the integration
of the radial velocity curve.}
%Both the distance and the average diameter
%of the star were adjusted simultaneously to the data.
\label{eta_aql_fit}
\end{figure}

\begin{figure}[t]
\centering
\includegraphics[bb=0 0 360 288, width=8.5cm]{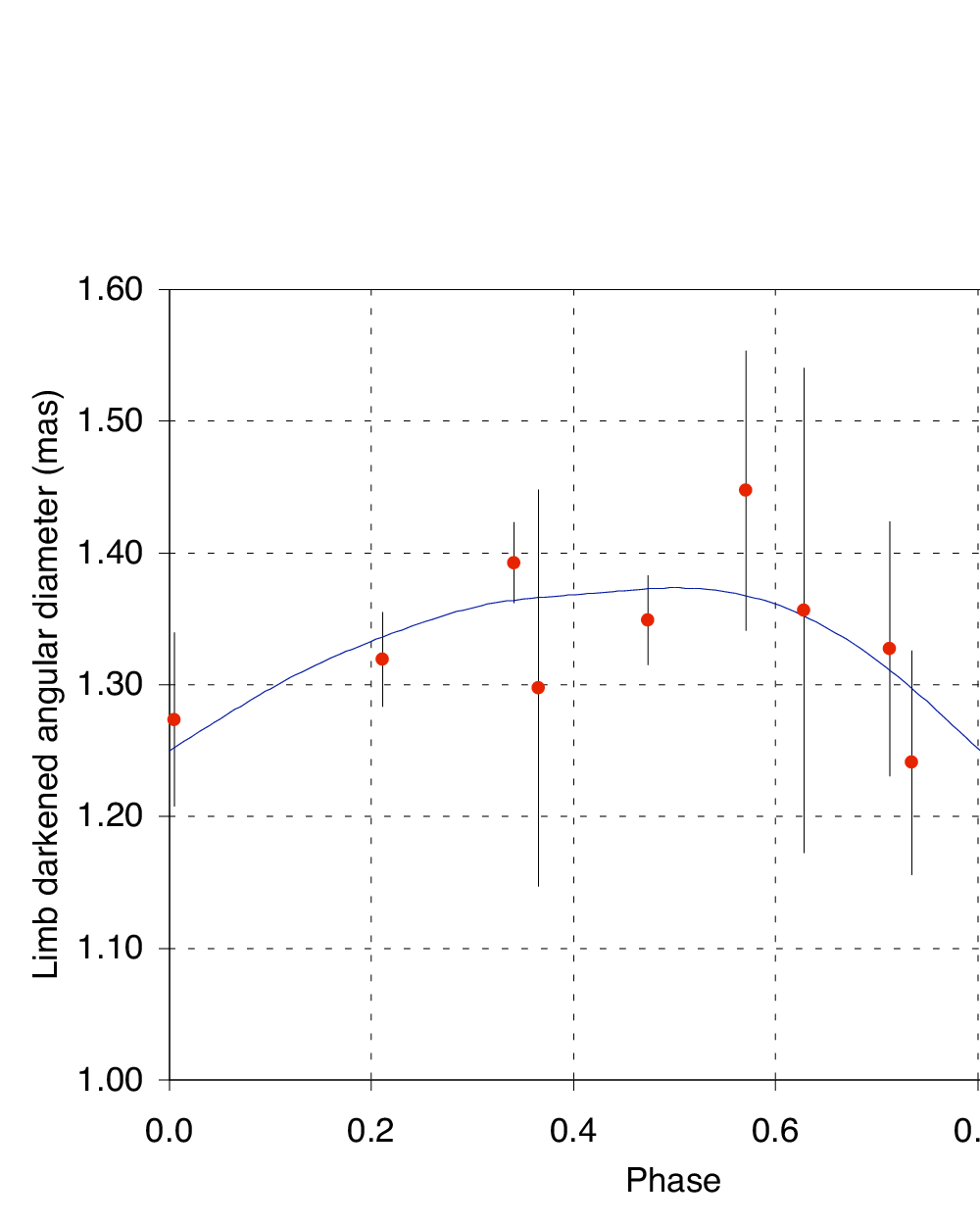}
\caption{Order 2 model fit for W\,Sgr.}
\label{w_sgr_fit}
\end{figure}

\begin{figure}[t]
\centering
\includegraphics[bb=0 0 360 288, width=8.5cm]{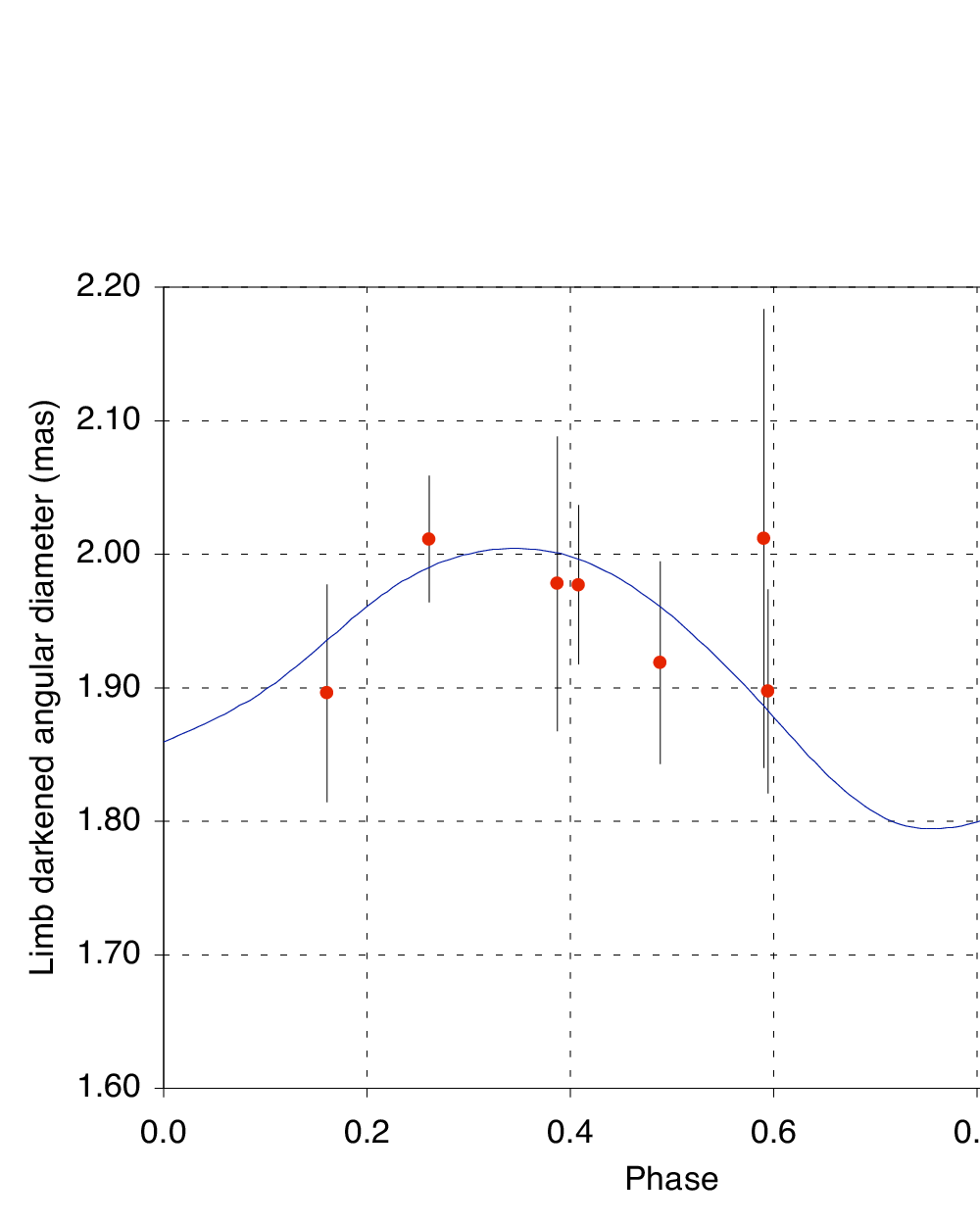}
\caption{Order 2 model fit for $\beta$\,Dor.}
\label{beta_dor_fit}
\end{figure}

\begin{figure}[t]
\centering
\includegraphics[bb=0 0 360 288, width=8.5cm]{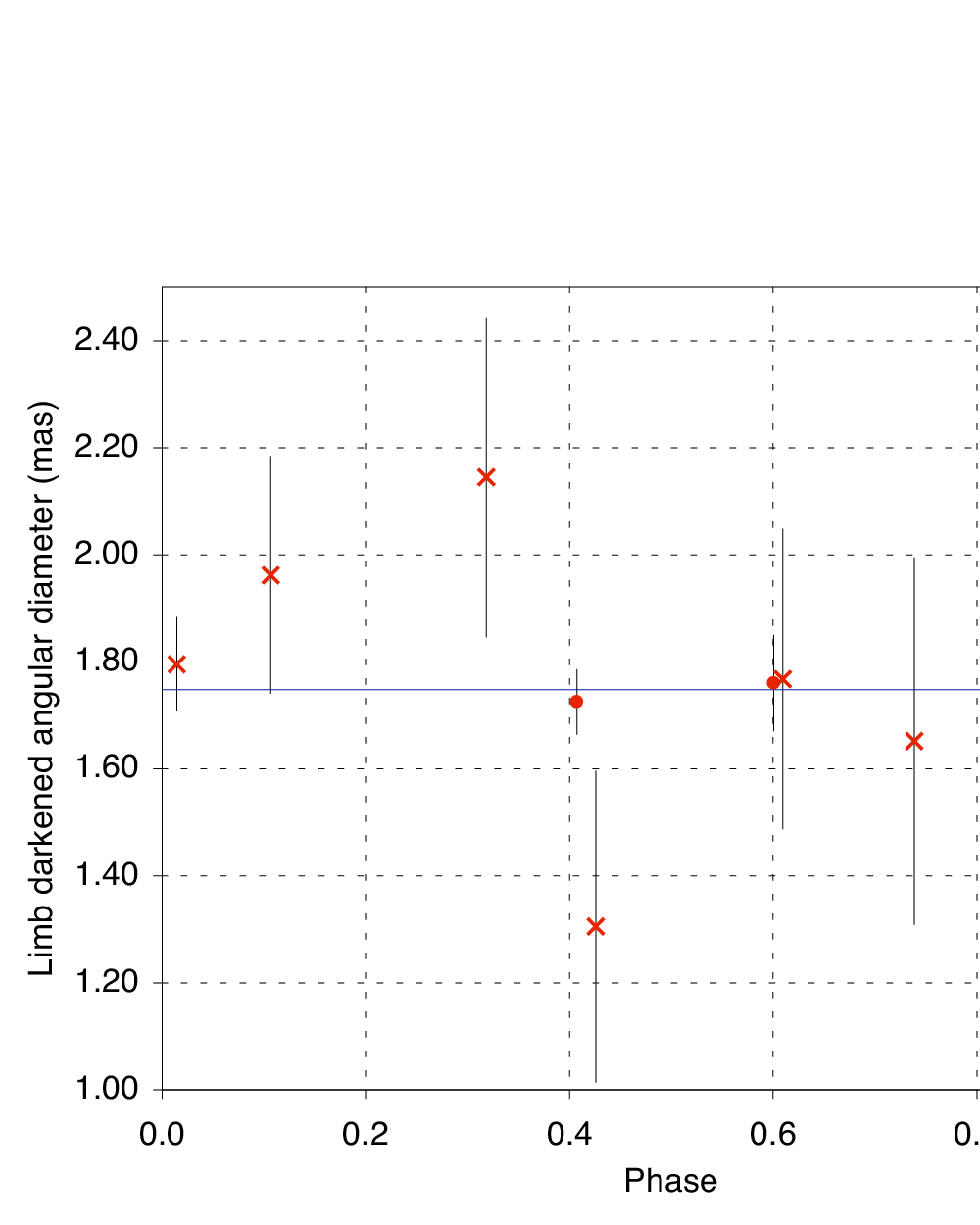}
\caption{Order 0 model fit for $\zeta$\,Gem. The crosses represent the
FLUOR/IOTA data, and the two points are UT1-UT3 observations with VINCI.}
\label{zeta_gem_fit}
\end{figure}

\begin{figure}[t]
\centering
\includegraphics[bb=0 0 360 288, width=8.5cm]{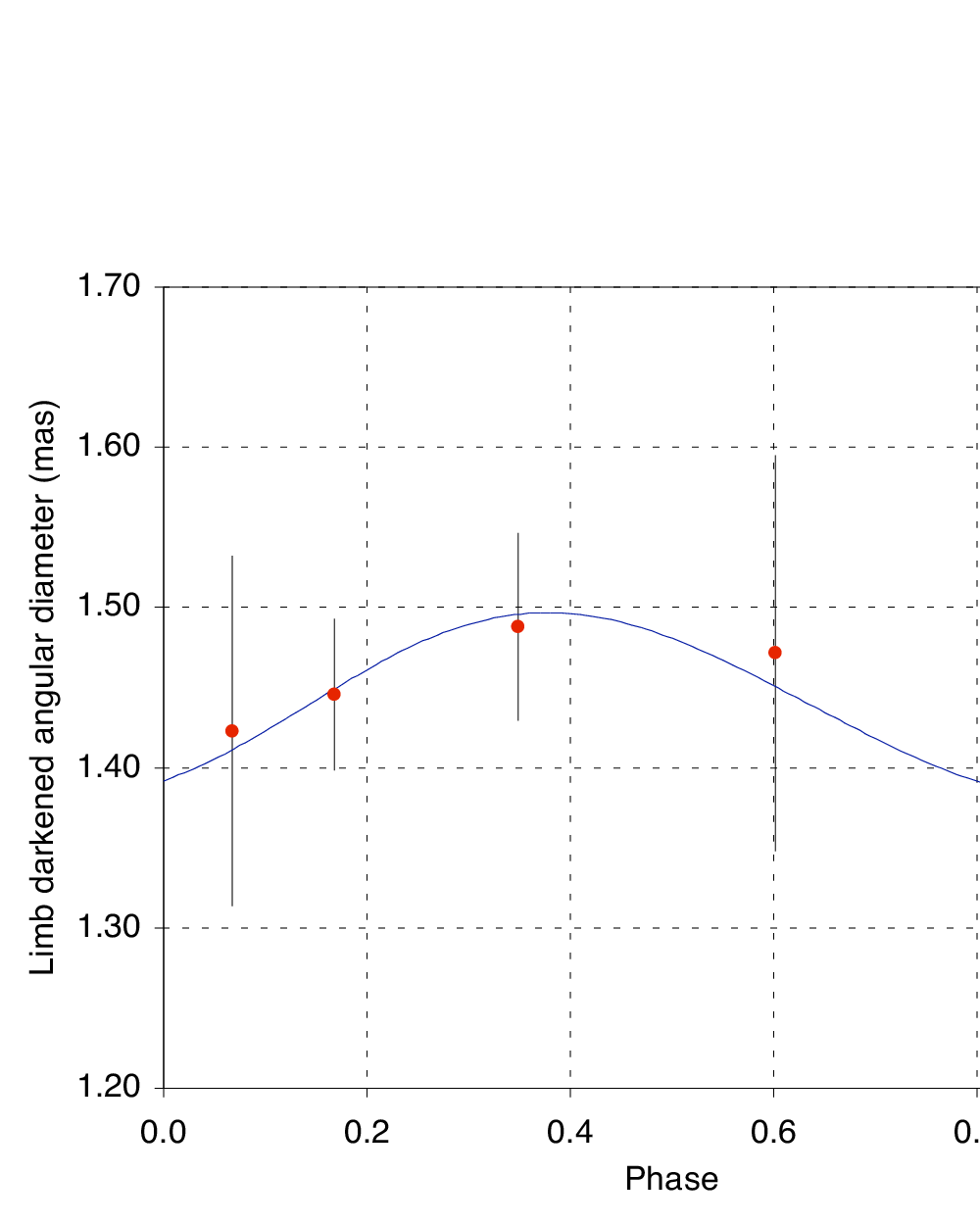}
\caption{Order 1 model fit for Y\,Oph.}
\label{y_oph_fit}
\end{figure}

\begin{figure}[t]
\centering
\includegraphics[bb=0 0 360 288, width=8.5cm]{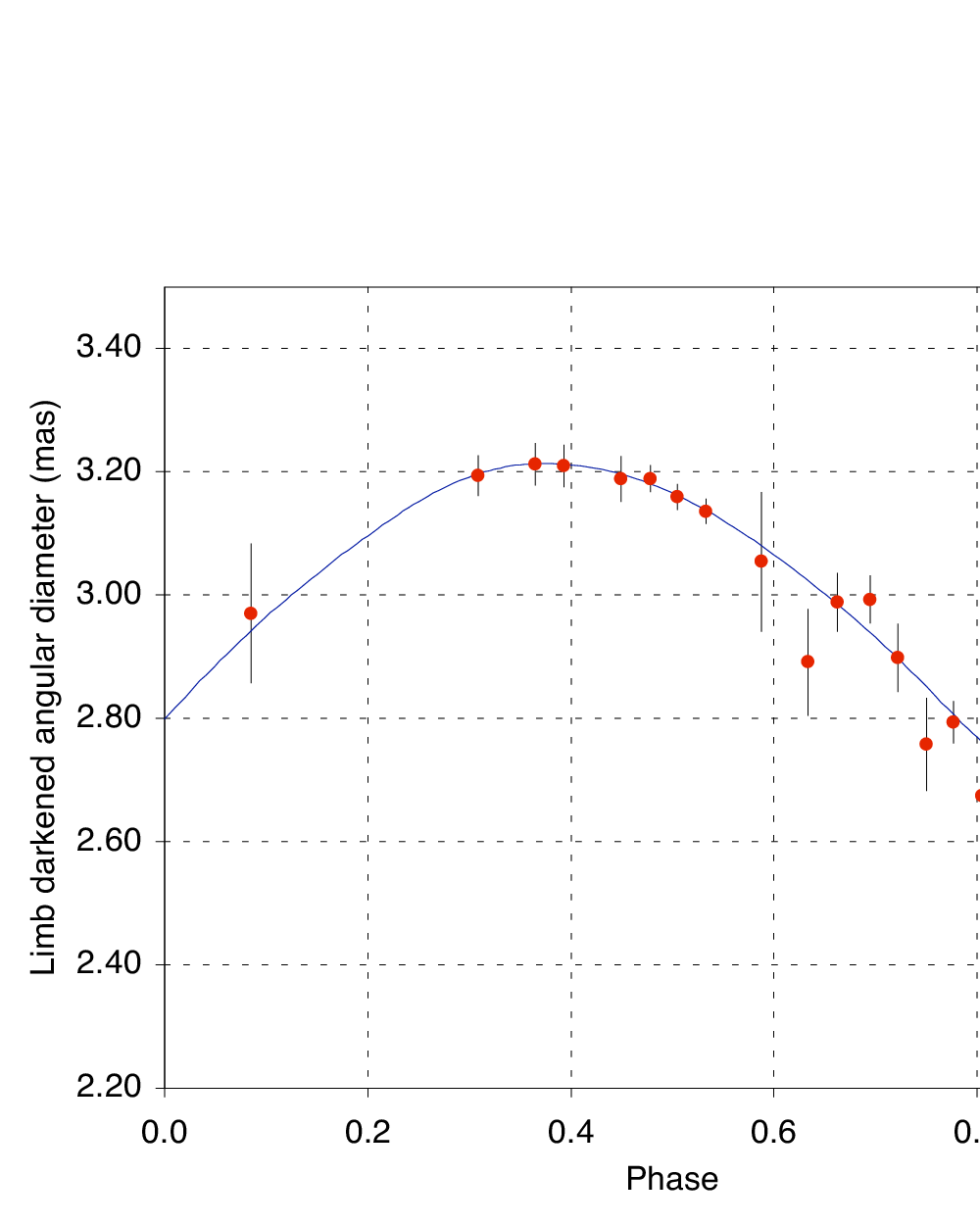}
\caption{Order 2 model fit for $\ell$\,Car.}
\label{l_car_fit}
\end{figure}

\begin{figure}[t]
\centering
\includegraphics[bb=0 0 360 288, width=8.5cm]{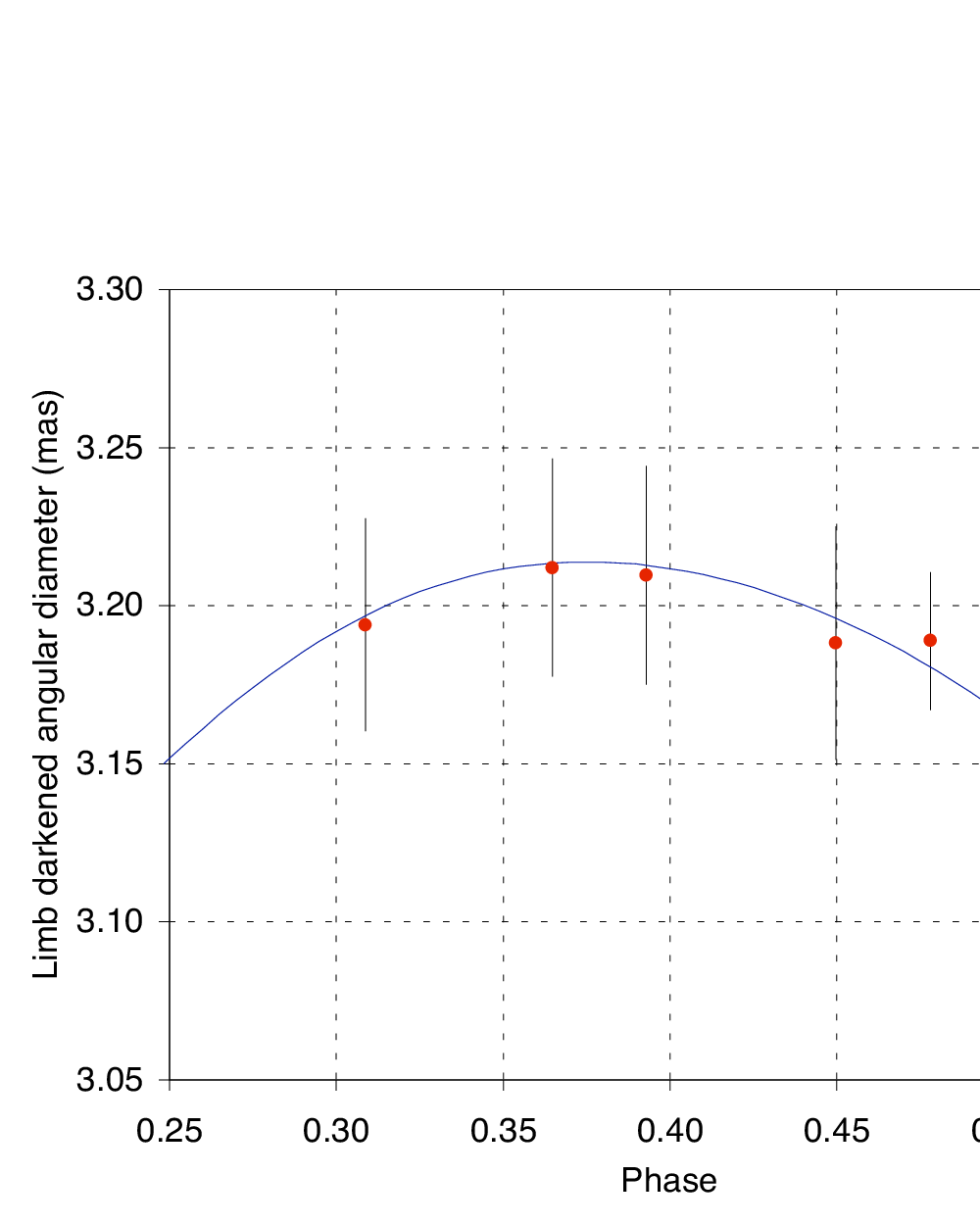}
\caption{Detail of Fig.\,\ref{l_car_fit} showing the angular diameter
curve of $\ell$\,Car around the maximum diameter.}
\label{l_car_fit_detail}
\end{figure}

Fig.~\ref{x_sgr_fit} to \ref{l_car_fit} show the best models for each star,
together with the VINCI/VLTI angular diameter measurements for the seven
Cepheids of our sample. Fig.\,\ref{l_car_fit_detail} gives an enlarged view of
the maximum diameter of $\ell$\,Car.

%______________ Figure
%______________ Figure
%______________ Figure
%______________ Figure
%______________ Figure
%______________ Figure
%______________ Figure
%______________ Figure

%___________________Table of fit results for constant diameter model, fit only on the distance
\begin{table}
\caption{{\bf Order 0.}
Cepheid average angular diameters and distances derived from the VINCI
interferometric measurements, assuming a constant diameter model ($\Delta D=0$),
The average diameter $\overline{D}$ is taken from Gieren et al.\,(\cite{gieren98}).
Two error bars are given in brackets for the angular diameter: the statistical dispersion
and the calibration systematics.
The uncertainty mentioned for the distance $d$ is the quadratic sum
of the statistical, calibration and P--R {\it a priori} diameter errors, the last two
being systematic in nature. The three types of errors are also reported
separately in brackets.
The results for $\ell$~Car
are mentioned only for completeness, but are not meant to be used
for further analysis, as our observations are inconsistent with a constant
diameter model.
}
\label{results_cepheids0}
\begin{tabular}{lclc}
\hline
\noalign{\smallskip}
Star & $\overline{\theta_{\rm LD 0}}$\,(mas) & $d_0$\,(pc) & $\chi^2_0$ \\
\noalign{\smallskip}
\hline
\noalign{\smallskip}
X\,Sgr & $1.471 \pm 0.033_{[0.013\ 0.031]}$ & $324 \pm 18_{[3\ 7\ 17]}$ & 0.38 \\
\noalign{\smallskip}
$\eta$~Aql & $1.856 \pm 0.028_{[0.009\ 0.026]}$ & $261 \pm 14_{[1\ 4\ 14]}$ & 3.98 \\
\noalign{\smallskip}
W\,Sgr & $1.348 \pm 0029_{[0.011\ 0.027]}$ & $376 \pm 22_{[3\ 8\ 21]}$ & 0.90\\
\noalign{\smallskip}
$\beta$\,Dor & $1.926 \pm 0.024_{[0.014\ 0.020]}$ & $319 \pm 20_{[3\ 2\ 19]}$ & 1.31 \\
\noalign{\smallskip}
$\zeta$\,Gem & $1.747 \pm 0.061_{[0.025\ 0.056]}$ & $360 \pm 25_{[5\ 12\ 22]}$ & 0.51\\
\noalign{\smallskip}
Y\,Oph & $1.459 \pm 0.040_{[0.023\ 0.033]}$ & $638 \pm 50_{[10\ 14\ 47]}$ & 0.16 \\
\noalign{\smallskip}
($\ell$\,Car) & $3.071 \pm 0.012_{[0.004\ 0.011]}$ & $524 \pm 49_{[1\ 2\ 49]}$ & 23.2 \\
\noalign{\smallskip}
\hline
\end{tabular}
\end{table}
%
%___________________Table of fit results for variable diameter model, only for distance
\begin{table}
\caption{{\bf Order 1.}
Cepheid angular diameters and distances, assuming the average diameter
$\overline{D}$ of Gieren et al. (\cite{gieren98}). The diameter variation curve
$\Delta D(\phi)$ is integrated from the radial velocity curve.
Only the distance is ajusted by the fitting procedure. The error bars on $d$ are
given as in Table~\ref{results_cepheids0}.}
\label{results_cepheids1}
\begin{tabular}{lclc}
\hline
\noalign{\smallskip}
Star & $\overline{\theta_{\rm LD 1}}$\,(mas) & $d_1$\,(pc) & $\chi^2_1$ \\
\noalign{\smallskip}
\hline
\noalign{\smallskip}
X\,Sgr & $1.461 \pm 0.033_{[0.013\ 0.031]}$ & $326 \pm 18_{[3\ 7\ 17]}$ & 1.36 \\
\noalign{\smallskip}
$\eta$~Aql & $1.839 \pm 0.028_{[0.009\ 0.026]}$ & $264 \pm 14_{[1\ 4\ 14]}$ & 0.40 \\
\noalign{\smallskip}
W\,Sgr & $1.312 \pm 0029_{[0.011\ 0.027]}$ & $386 \pm 22_{[3\ 8\ 21]}$ & 0.42\\
\noalign{\smallskip}
$\beta$\,Dor & $1.884 \pm 0.024_{[0.014\ 0.020]}$ & $326 \pm 20_{[3\ 2\ 19]}$ & 0.23 \\
\noalign{\smallskip}
$\zeta$\,Gem & $1.718 \pm 0.061_{[0.025\ 0.056]}$ & $366 \pm 25_{[5\ 12\ 22]}$ & 0.88 \\
\noalign{\smallskip}
Y\,Oph & $1.437 \pm 0.040_{[0.023\ 0.033]}$ & $648 \pm 51_{[10\ 15\ 47]}$ & 0.03 \\
\noalign{\smallskip}
$\ell$\,Car & $2.977 \pm 0.012_{[0.004\ 0.011]}$ & $542 \pm 49_{[1\ 2\ 49]}$ & 0.71 \\
\noalign{\smallskip}
\hline
\end{tabular}
\end{table}
%
%___________________Table of full fit
\begin{table}
\caption{{\bf Order 2.}
Cepheid average angular diameters and distances determined through the
application of the modified BW method. The only input is
the diameter variation curve $\Delta D(\phi)$ derived from the integration of the radial
velocity. The distance and average angular diameter are
ajusted simultaneously. The statistical and systematic errors on $d$ are
listed separately in brackets.}
\label{results_cepheids2}
\begin{tabular}{lclc}
\hline
\noalign{\smallskip}
Star & $\overline{\theta_{\rm LD 2}}$\,(mas) & $d_2$\,(pc) & $\chi^2_2$ \\
\noalign{\smallskip}
\hline
\noalign{\smallskip}
$\eta$\,Aql & $1.839 \pm 0.028_{[0.009\ 0.026]}$ & $276^{+55}_{-38}\ {[^{55\ 6}_{38\ 4}]}$ & 0.43 \\
\noalign{\smallskip}
W\,Sgr & $1.312 \pm 0.029_{[0.011\ 0.027]}$ & $379^{+216}_{-130}\ {[^{216\ 11}_{130\ 7}]}$ & 0.48 \\
\noalign{\smallskip}
$\beta$\,Dor & $1.891 \pm 0.024_{[0.014\ 0.020]}$ & $345^{+175}_{-80}\ {[^{175\ 5}_{80\ 2}]}$ & 0.25 \\
\noalign{\smallskip}
$\ell$\,Car & $2.988 \pm0.012_{[0.004\ 0.011]}$ & $603^{+24}_{-19}\ {[^{24\ 3}_{19\ 2}]}$ & 0.49 \\
\noalign{\smallskip}
\hline
\end{tabular}
\end{table}
%
%
%
%_______________________________________________________ Section
%
\section{Discussion}

\subsection{Limb darkening of $\eta$\,Aql and $\zeta$\,Gem}

From the NPOI (Armstrong et al.~\cite{armstrong01}, Nordgren et al.~\cite{nordgren00}),
PTI (Lane et al.~\cite{lane02}) and VINCI/VLTI measurements, we know the
average UD angular diameters of $\eta$\,Aql and $\zeta$\,Gem
at several effective wavelengths with high precision.
Table~\ref{ea_zg_UD} gives the angular diameter values and the corresponding wavelengths.
Claret's~(\cite{claret00}) linear limb darkening parameters $u$ were used to
compute the expected conversion factors $\rho = \theta_{\rm LD} / \theta_{\rm UD}$.
To read the $u$ table, we have considered the closest parameters to the average
values for $\eta$\,Aql and $\zeta$\,Gem in Table~\ref{cepheids_params},
and we computed $\rho$ using the formula from Hanbury Brown el al.~(\cite{hanbury74}):
\begin{equation}
\rho = \sqrt{\frac{1-u/3}{1-7u/15}}
\end{equation}
For the NPOI observation ($\lambda_{\rm eff} \approx 0.73\,\mu$m),
we have chosen an intermediate value of $u$ between the $R$ and $I$ bands.

We note that the value of $\theta_{\rm LD}$ for $\eta$\,Aql that we derive for the
NPOI observation, $\theta_{\rm LD} = 1.73 \pm 0.04$\,mas, is not identical to
the LD angular diameter originally given by Armstrong et al.\,(\cite{armstrong01}),
$\theta_{\rm LD} = 1.69 \pm 0.04$\,mas. There is a 1$\sigma$ difference, that may
be due to the different source of limb darkening coefficient that these authors
used for their modeling (Van Hamme~\cite{vanhamme93}).

The resulting $\theta_{\rm LD}$ values for the three observations are compatible
at the 2\,$\sigma$ level, but there is a slight trend that points towards an
underestimation of the limb darkening effect at shorter wavelengths,
or alternatively its overestimation at longer wavelengths.
Considering that the limb darkening is already small in the infrared,
the first hypothesis seems more plausible.
Marengo et al.~(\cite{marengo02}, \cite{marengo03}) have
shown that the Cepheids limb darkening can be significantly
different from stable giant stars, particularly at visible wavelengths.
This could explain the observed difference between the
0.73\,$\mu$m and $K$ band diameters of $\eta$\,Aql and $\zeta$\,Gem,
the latter being probably closer to the true LD diameters,
thanks to the lower limb darkening in the infrared.

In the case of $\eta$\,Aql, another explanation
could be that the measurement at visible wavelengths
is biased by the blue companion of $\eta$\,Aql. However, it is 4.6
magnitudes fainter than the Cepheid in the $V$ band (B\"ohm-Vitense \&
Proffitt~\cite{bohm85}, see also Sect.\,\ref{binarity}),
and therefore should not contribute significantly to the visibility of the fringes.
%
%___________________Table Eta Aql Zeta Gem UD diams
\begin{table}
\caption{Average UD angular diameter of $\eta$\,Aql and $\zeta$\,Gem
from the litterature, and the associated conversion factor
$\rho = \theta_{\rm LD} / \theta_{\rm UD}$ from the linear limb darkening
coefficients of Claret\,(\cite{claret00}). References:
(1) Armstrong et al.\,(\cite{armstrong01})
and Nordgren et al.\,(\cite{nordgren00}),
(2) Lane et al.\,(\cite{lane02}),
(3) this work.}
\label{ea_zg_UD}
\begin{tabular}{lcccc}
\hline
\noalign{\smallskip}
Ref. & $\lambda$\,($\mu$m) & $\theta_{\rm UD}$\,(mas) & $\rho$ & $\theta_{\rm LD}$\,(mas) \\
\hline
\noalign{\smallskip}
$\eta$\,Aql\\
(1) & 0.73 & $1.65 \pm 0.04$ & 1.048 & $1.73 \pm 0.04$ \\
(2) & 1.65 & $1.73 \pm 0.07$ & 1.024 & $1.77 \pm 0.07$ \\
(3) & 2.18 & $1.80 \pm 0.03$ & 1.021 & $1.84 \pm 0.03$\\
\hline
\noalign{\smallskip}
$\zeta$\,Gem\\
(1) & 0.73 & $1.48 \pm 0.08$ & 1.051 & $1.56 \pm 0.08$\\
(2) & 1.65 & $1.61 \pm 0.03$ & 1.027 & $1.65 \pm 0.03$\\
(3) & 2.18 & $1.70 \pm 0.06$ & 1.023 & $1.75 \pm 0.06$\\
\noalign{\smallskip}
\hline
\end{tabular}
\end{table}

\subsection{Binarity and other effects \label{binarity}}

As demonstrated by several authors
(see Szabados~\cite{szabados03} for a complete database),
binarity and multiplicity are common in the Cepheid class.
Evans~(\cite{evans92a}) has observed that 29\,\% of the
Cepheids of her sample have detectable companions.

Our sample of Cepheids contains four confirmed binary Cepheids, out
of a total of seven stars.
As it is biased towards bright and nearby Cepheids, 
this large fraction is an indication that many Cepheids currently believed to be
single could have undetected companions.
X\,Sgr (Szabados~\cite{szabados89b}),
$\eta$\,Aql (B\"ohm-Vitense \& Proffitt~\cite{bohm85}),
and W\,Sgr (B\"ohm-Vitense \& Proffitt~\cite{bohm85}, Babel et al.\,\cite{babel89})
are confirmed members of binary or multiple systems. 
$\zeta$\,Gem is a visual binary star (Proust et al.~\cite{proust81}), but the
separated companion does not contribute to our observations.
Y\,Oph was once suspected to be a binary (Pel~\cite{pel78}),
but Evans~(\cite{evans92a}) has not confirmed the companion, and has
set an upper limit of A0 on its spectral type.

The physical parameters of the companions of $\eta$\,Aql and W\,Sgr
have been derived by B\"ohm-Vitense \& Proffitt~(\cite{bohm85})
and Evans~(\cite{evans91}), based on ultraviolet spectra.
The latter has derived spectral types of B9.8V and A0V, respectively.
The orbital parameters of the binary W\,Sgr were computed by
Babel et al.\,(\cite{babel89}) and Albrow \& Cottrell\,(\cite{albrow96}).
Based on IUE spectra, Evans~(\cite{evans92a}) has
set an upper limit of A0 on the spectral type of the companion of X\,Sgr.

The difference in $V$ magnitude between these three Cepheids and their
companions is $\Delta {\rm M}_V \ge 4.5$.
% Mv A0 ~ 0.3
The $\Delta {\rm M}_K$ is even larger due to the blue color of these stars,
$\Delta {\rm M}_K \ge 5.7$.
Therefore, the effect on our visibility measurements is negligible, with
a potential bias of $\Delta V^2 \le 0.5\,\%$. For example, this translates into a
maximum error of $\pm 11\,\mu$as on the average angular diameter of $\eta$\,Aql,
(a relative error of $\pm 0.6\,\%$), that is significantly smaller than our
error bars ($\pm 1.5\,\%$).
In the $K$ band, the effect of the companions of the other
Cepheids is also negligible at the precision level of our measurements.
However, the presence of companions will have
to be considered for future measurements with angular diameter precisions
of a few $\mu$as. In this respect, long-period Cepheids, such
as $\ell$\,Car, are more reliable, as their intrinsic brightness is larger than
the short-period pulsators, and therefore they dominate their potential
companions even more strongly.

Fernie et al.~(\cite{fernie95b}) have found that the amplitude of the
light curve of Y\,Oph has been decreasing for a few decades. A similar behavior
has been observed only on Polaris (e.g. Evans et al.~\cite{evans02}).
The uncertainty on our $\theta_{\rm LD}$ measurements has not allowed us to
detect unambiguously the pulsation of this star, but it is clearly an
important target for future observations using the Auxiliary Telescopes
(1.8\,m) of the VLTI in order to estimate its parameters with high precision.

Interestingly, Gieren et al.~(\cite{gieren93}) have studied
the impact of binary Cepheids on their determination of the
period-luminosity relation using 100 Cepheids,
and they conclude that it is negligible. This is due to the very large
intrinsic luminosity of the Cepheids that overshine by several orders of magnitude
most of the other types of stars.
%
%_______________________________________________________ Section
%
\section{Conclusion and perspectives}
We have reported in this paper our long-baseline interferometric
observations of seven classical Cepheids using the VINCI/VLTI instrument.
For four stars ($\eta$\,Aql, W\,Sgr, $\beta$\,Dor and $\ell$\,Car),
we were able to apply a modified version of the BW method, resulting in an
independent estimate of their distance.
For all stars, we also derived their distances from lower order
fitting methods, that use an {\it a priori} estimate of their linear diameter from
the P--R relation of Gieren et al.~(\cite{gieren98}).
We would like to emphasize that the order 0/1 and order 2
error bars are different in nature, and they should be
treated differently in any further use of these results.
While the order 2 error bars can be treated as statistical
(i.e. reduced by averaging), the order 0/1 methods errors are
dominated by the systematic uncertainty introduced by the {\it a priori}
estimation of the linear radius.
The respective contributions of the statistical and systematic uncertainties
are given separately in Tables~\ref{results_cepheids0} and \ref{results_cepheids1}.
These values assume a constant value of the $p$-factor of 1.36, and can
be scaled linearly for other values.

We will use these distances in Paper II, together with previously published measurements,
to calibrate the zero points of the Period-Radius and Period-Luminosity relations
%to a relative precision better than 2 and 5\,\%, respectively, and we find a 
%satisfactory agreement with previously published work.
In Paper III, we will calibrate the surface brightness--color relation, with a particular emphasis on the
evolution of $\ell$\,Car in this diagram over its pulsation. These three empirical
relations are of critical importance for the extragalactic distance scale.

The direct measurement of the limb darkening of nearby Cepheids by interferometry
is the next step of the interferometric study of these stars. It will allow a refined
modeling of the atmosphere of these stars. This observation will be achieved soon
using in particular the long baselines of the VLTI equipped with the AMBER instrument,
and the CHARA array for the northern Cepheids.
Another improvement of the interferometric BW methow will come from
radial velocity measurements in the near infrared (see e.g. Butler \& Bell~\cite{butler97}).
They will avoid any differential limb darkening between the interferometric
and radial velocity measurements, and therefore make the resulting distances
more immune to limb darkening uncertainties.
%
%The measured distances will allow us in a forthcoming paper to establish a new calibration
%of the Cepheids P--L relation, as well as the P--R and surface brightness--color relations.

%___________________Table of fit results
%\begin{table}
%\caption{Summary of the best fit LD angular diameters and distances 
%of the seven Cepheids observed with VINCI. The detailed results for each
%star and fitting order are presented in Tables \ref{results_cepheids0}
%to \ref{results_cepheids2}. }
%\label{results_cepheids_summary}
%\begin{tabular}{lccll}
%\hline
%\noalign{\smallskip}
%Star & Order & $\overline{\theta_{\rm LD}}$\,(mas) & $d$\,(pc) \\
%\noalign{\smallskip}
%\hline
%\noalign{\smallskip}
%X~Sgr & 0 & $1.471 \pm 0.033$ & $324 \pm 18$ \\
%\noalign{\smallskip}
%$\eta$~Aql & 2 & $1.839 \pm 0.028$ & $276^{+55}_{-38}$ \\
%\noalign{\smallskip}
%W~Sgr & 2 & $1.312 \pm 0.029$ & $379^{+216}_{-130}$\\
%\noalign{\smallskip}
%$\beta$~Dor & 2 & $1.891 \pm 0.024$ & $345^{+175}_{-80}$ \\
%\noalign{\smallskip}
%$\zeta$~Gem & 0 & $1.747 \pm 0.061$ & $360 \pm 25$\\
%\noalign{\smallskip}
%Y~Oph & 1 & $1.438 \pm 0.051$ & $648 \pm 52$ \\
%\noalign{\smallskip}
%$\ell$~Car & 2 & $2.988 \pm 0.012$ & $603^{+24}_{-19}$ \\
%\noalign{\smallskip}
%\hline
%\end{tabular}
%\end{table}

\begin{acknowledgements}
DB acknowledges support from NSF grant AST-9979812. PK acknowledges
support from the European Southern Observatory through a postdoctoral fellowship.
Based on observations collected at the European Southern Observatory,
Cerro Paranal, Chile, in the framework of ESO shared-risk programme
071.D-0425 and unreferenced commissioning programme in P70.
The VINCI/VLTI public commissioning data reported in this paper
have been retrieved from the ESO/ST-ECF Archive (Garching, Germany).
This work has made use of the wavelet data processing technique,
developed by D. S\'egransan (Observatoire de Gen\`eve),
and embedded in the VINCI pipeline.
This research has made use of the SIMBAD database at CDS, Strasbourg (France).
We are grateful to the ESO VLTI team, without whose efforts no observation
would have been possible.
\end{acknowledgements}

{}
\end{document}